\documentclass[%
  reprint,
  onecolumn,
  superscriptaddress,
  amsmath,amssymb,
  aps,
  11pt,
  pre,
  longbibliography
]{revtex4-1}

\usepackage{geometry}
\usepackage{tgtermes}

\usepackage{graphicx}   % need for figures
\usepackage{epstopdf}
\usepackage{soul}
\usepackage{hyperref}   % use for hypertext links, including those to external 
\usepackage{bm}
\usepackage{amsfonts}
\usepackage{natbib}

\usepackage{color}      % use if color is used in text
\DeclareMathOperator{\sign}{sign}
\usepackage[normalem]{ulem} %for barred text
\usepackage{soul}	%for correctly underlined text

\def\ukp{u^+_{\bm k}}
\def\ukm{u^-_{\bm k}}

\def\ukpm{u^\pm_{\bm k}}

\def\bk{{\bm k}}

\def\br{{\bm r}}
\def\bq{{\bm q}}
\def\bu{{\bm u}}
\def\bfo{{\bm f}}
\def\bx{{\bm x}}
\def\bz{{\bm z}}
\def\bw{{\bm \omega}}
\def\hm{{\bm h}^-_{\bm k}}
\def\hp{{\bm h}^+_{\bm k}}
\def\hpm{{\bm h}^\pm_{\bm k}}

\begin{document}
\title{Helicity statistics in homogeneous and isotropic turbulence and 
turbulence models \footnote{Version accepted for publication (postprint) on  Phys. Rev. Fluids 2, 024601 -- Published 1 February 2017}}
\author{Ganapati Sahoo}
\affiliation{Department of Physics \& INFN, University of Rome Tor Vergata, Via 
della Ricerca Scientifica 1, 00133 Rome, Italy.}
\author{Massimo De Pietro} 
\affiliation{Department of Physics \& INFN, University of Rome Tor Vergata, Via 
della Ricerca Scientifica 1, 00133 Rome, Italy.}
\author{Luca Biferale} 
\affiliation{Department of Physics \& INFN, University of Rome Tor Vergata, Via 
della Ricerca Scientifica 1, 00133 Rome, Italy.}
\date{\today}

\begin{abstract}
 
We study  the
statistical properties of helicity in direct numerical simulations of fully
developed homogeneous and isotropic turbulence and in a class of turbulence
shell models. We consider correlation functions based on combinations of
vorticity and velocity increments that are not invariant under mirror symmetry.
We also study the scaling properties of high-order structure functions based on
the moments of the  velocity increments projected on a subset of modes with
either positive or negative helicity (chirality). We show that mirror symmetry
is recovered at small-scales, i.e., chiral terms are subleading and they
are well captured by a dimensional argument plus anomalous corrections.
These findings are
also supported by a high Reynolds numbers study of helical shell models with the
same chiral symmetry of Navier-Stokes equations.
\end{abstract}

\maketitle

\section{Introduction}

All phenomenological theories of three dimensional (3D) turbulence are based on
the concept of direct energy cascade \cite{frisch1995turbulence}.  However,
helicity is also an inviscid invariant of the 3D Navier-Stokes equations (NSEs)
defined as the scalar product of velocity $\bu(\bx)$ with vorticity
$\bw(\bx)$. Its mean value 
\begin{align}  
H  = \frac{1}{V}\int_V d^3 x \, \bu(\bx) \cdot \bw(\bx) \, , 
\end{align} 
is exactly zero if the flow is invariant under mirror symmetry, $\bw$ being a
pseudovector.  Since its discovery \cite{moffatt69,moffatt92,brissaud1973},
helicity has been the object of many speculations. In particular, it is not
clear if the presence of a nonzero mean helicity, globally or locally, can
affect the statistical properties of the forward energy cascade. On the one hand,
because the nonlinear term of the NSE is locally proportional to the solenoidal
component of $\bu \times \bw$, flows
with a nonzero helicity might have a  strongly depleted energy transfer
\cite{kraichnan1988depression,moffatt2014helicity}. On the other hand, helicity
is not sign-definite, and therefore cancellations might eventually smooth-down
this {\it blocking} mechanism \cite{chen2003joint,chen2003prl}. There
exist instances  where helicity plays a key role, interfering with
the energy transfer, as  in rotating turbulence
\cite{mininni2009,deusebio2014}, in  shear flows \cite{dubrulle2014}, and in
the case of an NSE confined to evolve on a subset of sign-definite helical modes
\cite{biferale2012,biferale2013,sahoo2015,sahooEPJE}. In the presence of a
stationary helicity injection, we have  an exact law which predicts the
scaling properties of a specific velocity-vorticity mixed  third-order
correlation function \cite{GomezPolitano,kurien_isotropic,gledzer2015inverse}.
Nevertheless, this is not a strong constraint for the whole statistics. Indeed,
different phenomenological scaling for the  spectral properties
has been proposed in the presence of two simultaneous fluxes of energy and
helicity \cite{brissaud1973,chen2003joint,kurien_cascade}.

In this paper we further investigate the statistical properties of
helicity in fully developed turbulence by using high resolution direct
numerical simulations (DNSs).  In order to have a proper way to distinguish the
importance of mirror-symmetry-breaking contributions {\it scale-by-scale}, we
study the  properties of a class of structure functions based on velocity
increments decomposed on positive or negative helical modes. The latter have
the advantage of observable definitions that are sensitive to lack of mirror
symmetry for all moments, odd or even,  differently from what was proposed earlier
in Refs. \cite{chen2003prl,mininni_pouquet}. Furthermore, we also introduce a set of
velocity-vorticity correlation functions based on the helicity cancellation
exponent \cite{cancellation} that allows us to quantify the breaking of mirror
symmetry also on quantities based on velocity gradients.

We show that helicity-sensitive observables  are always subleading with respect 
to the ones dominated by the energy flux. Results are also supported  
by studying analogous quantities in a helical shell model 
\cite{biferale2003shell, biferale_helical_1996, chen2003joint}. We show that 
the scaling behavior of chiral quantities is well captured by an analytical  
contribution in terms of the helicity flux, plus  a small anomalous correction.

\section{Phenomenological background \label{sec:section2}}

We consider the  3D forced NSE: 
\begin{equation} 
  \label{eq:ns+++}
  \partial_t \bu + \bu \cdot {\bm \nabla} \bu=  -{\bm \nabla} p   +\nu \Delta 
  \bu + \bfo \, ,
\end{equation}
where $p$ is the pressure, $\nu$ is the kinematic viscosity, and $\bfo$ is a 
parity-breaking external forcing mechanism with energy injection, 
$\varepsilon = \langle \bu \cdot \bfo \rangle$ and helicity injection,  $h 
= \langle \bu \cdot ({\bm \nabla} \times \bfo) +  \bw \cdot \bfo 
\rangle$. Under the assumptions of stationarity, homogeneity, and 
isotropy (but not mirror-symmetry) it is possible to derive two exact 
equations for two-point third-order correlation functions 
\cite{frisch1995turbulence, GomezPolitano, kurien_isotropic,Chkhetiani1996}: 
\begin{align}
&  \langle (\delta_r u)^3\rangle = -\frac{4}{5} \varepsilon\, r \, , 
\label{eq:45}\\
&  \langle \delta_r u (\delta_r \bm u \cdot \delta_r \bw)\rangle - 
\frac{1}{2}\langle \delta_r \omega (\delta_r \bm u \cdot \delta_r \bm 
u)\rangle  = -\frac{4}{3} h\, r \, , \label{eq:45_second}
\end{align}
where $\delta_r u$ and $\delta_r \omega$ are, respectively, the longitudinal
velocity and vorticity increments, defined in terms of the projection on the
unit  vector $\hat \br$: $\delta_r X = \delta_r \bm X \cdot \hat \br$, 
and the generic vector increment between two points is $\delta_r \bm X = \bm
X(\br +\bx ) -\bm X(\bx)$. Notice that (\ref{eq:45_second}) is different from
zero only in the presence of a mirror-breaking forcing mechanism.  The two exact
scaling relations (\ref{eq:45})--(\ref{eq:45_second}) are valid in the inertial
range, i.e., when the increment $r$ is chosen in a range of scales where
dissipative and forcing effects can be neglected. Moreover, since helicity is
not sign-definite, it is not possible to  predict the energy transfer 
direction: both a 
simultaneous cascade of energy and helicity toward
small-scales and a split cascade with energy flowing upward and helicity 
downward are
possible \cite{brissaud1973,kraichnan1971,biferale2013,dubrulle2014}. In order
to disentangle in a systematic way the statistical properties under mirror
symmetry, it is useful to adopt an exact decomposition of the velocity field in
positive and negative Fourier helical waves \cite{constantin,waleffe}: 
\begin{align}
  \bu(\bx,t) = \sum_{\bk}  [\ukp(t) \hp  + \ukm(t) \hm] \mathrm{e}^{i \bk \cdot 
  \bx} \, ,
\end{align}
where $\hpm$ are the eigenvectors of the curl, i.e., $i {\bk} \times 
\hpm = \pm k \hpm$. We choose $\hpm = \hat{\mu}_{\bm k} \times \hat{k} 
\pm i \hat{\mu}_{\bm k}$, where $ \hat{\mu}_{\bm k}$ is a unit vector 
orthogonal to ${\bk}$ satisfying the condition $\hat{\mu}_{\bm k} = - 
\hat{\mu}_{-\bk}$, e.g., $\hat{\mu}_{\bk} = {\bz} \times {\bk} /|| 
{\bz} \times {\bk} ||$, with any arbitrary vector ${\bz}$. In terms of 
such  decomposition the total energy, $E = \int d^3 x \, |\bu(\bx)|^2 
$, and the total helicity are written as
\begin{align}
&    E = \sum_{\bk} |\ukp|^2 + |\ukm|^2 \, , \label{eq:Etot}\\
&  H = \sum_{\bk} k\,(|\ukp|^2 - |\ukm|^2) \,. \label{eq:Htot}
\end{align} 
It is useful to further distinguish the
energy content of the positive and negative helical modes, $E^\pm(k) =  \sum_{k
\le |\bk| < k+1} |\ukpm|^2$, such that we have for  the energy and helicity 
spectra
\cite{chen2003joint}: 
\begin{align}
&    E(k)  = E^+(k) + E^-(k) \, , \label{eq:epm}\\
&    H(k)  = k\,[E^+(k) - E^-(k)] \, . \label{eq:epm_2}
\end{align} 
It is straightforward to realize that the equivalent of 
(\ref{eq:epm})-(\ref{eq:epm_2})  
in real space is given by the second-order correlation functions 
decomposed  in terms of the fields $\bu^\pm (\bx) =  \sum_{\bk}  
\ukpm(t) \hpm \exp^{i \bk \cdot \bx}$ :
\begin{align}
\label{eq:epm2}
&   \langle \delta_r u_i \delta_r u_i\rangle   =  \langle \delta_r u^+_i 
\delta_r u^+_i\rangle   + \langle \delta_r u^-_i \delta_r u^-_i\rangle \, , \\
&   \langle \delta_r u_i \delta_r \omega_i\rangle   =  \langle \delta_r u^+_i 
\delta_r \omega^+_i\rangle   
+  \langle \delta_r u^-_i \delta_r \omega^-_i\rangle \, ,
\end{align} 
because both  mixed terms
$\langle \delta_r u^\pm_i \delta_r \omega^\mp_i \rangle$ and $\langle \delta_r
u^\pm_i \delta_r u^\mp_i \rangle$  vanish, due to the orthonormality of $\hpm$.
\\ It is not possible to derive
a closed expression for the energy and helicity spectra from
(\ref{eq:45})-(\ref{eq:45_second}) alone, because there exists a continuum of
possible combinations of $\varepsilon, h$ and $k$ with the correct dimensional
properties:
\begin{align}
E(k) = \varepsilon^{\frac{2}{3}-\alpha} h^\alpha k^{-\frac{5}{3}-\alpha} \, .
\end{align} 
Different possibilities have been proposed, based on different closures of 
the spectral equations, depending on the dynamical time-scale 
that drives the energy and helicity transfers. One possibility is  
based on the idea that the  only relevant time-scale is the 
one given by the energy fluctuations, $\tau^E_r \sim 
\varepsilon^{-1/3} r^{2/3}$. In this case we have the dimensional  
estimate for the (mirror invariant) energy flux:
\begin{align}
\varepsilon \sim \langle \delta_r u_i \delta_r u_i\rangle /\tau_r^E \rightarrow 
\langle (\delta_r u)^2\rangle  \sim \varepsilon^{2/3}r^{2/3} \, ,
\label{eq:fluxe}
\end{align}
while for the chiral  term,
\begin{align}
h  \sim \langle \delta_r u_i \delta_r \omega_i \rangle /\tau_r^E \rightarrow 
\langle \delta_r u_i \delta_r \omega_i \rangle  \sim h 
\varepsilon^{-1/3}r^{2/3} \, .
\label{eq:fluxh}
\end{align}
Translating back to Fourier space we would then have for the semi-sum 
(mirror-symmetric) and the semi-difference (mirror-antisymmetric)  
of the spectral components \cite{chen2003joint}:
\begin{align}
&   E^+(k) +   E^-(k)  \sim C_E \varepsilon^{2/3} k^{-5/3} \, 
,\label{eq:chen1}\\
&  E^+(k) - E^-(k) \sim C_H h \varepsilon^{-1/3} k^{-8/3}  \, ,\label{eq:chen2}
\end{align} 
where $C_E$ and $C_H$ are two dimensionless constants. Hence, the two energy 
components can be written as: 
\begin{align}
E^\pm(k)  \sim C_E\varepsilon^{2/3} k^{-5/3} \pm C_H h \varepsilon^{-1/3} 
k^{-8/3}  \, . \label{eq:epm-chen}
\end{align} 
Another possible dimensional closure  employs the   helicity 
time-scale, $\tau_r^H \sim h^{-1/3} r^{1/3}$, to  evaluate both  fluxes 
(\ref{eq:fluxe})--(\ref{eq:fluxh}). In this case we have 
\cite{kurien_cascade}: 
\begin{align}
E^\pm(k)  \sim C_E \varepsilon h^{-1/3} k^{-4/3} \pm C_H h^{2/3}  k^{-7/3}  \, 
. \label{eq:epm-kurien}
\end{align}
Relation (\ref{eq:epm-kurien}) breaks the $-5/3$ law for the energy spectrum 
and has
been proposed to be valid only in the high-$k$ region of strongly helical
turbulence, to explain the bottleneck observed close to the viscous scale.
Indeed, relation  (\ref{eq:epm-kurien}) is not smooth for $h
\rightarrow 0$ and therefore cannot be considered a good option if helicity is
subleading. 
A third possible scenario is a split cascade, 
where energy flows upward and helicity downward. In this case, in 
the forward-helicity cascade range, only $h$ flux appears, and the 
dimensional prediction gives:
\begin{align}
E(k) \sim h^{2/3} k^{-7/3} \, , \qquad H(k) \sim h^{2/3} k^{-4/3}. 
\end{align}
This last scenario has never been observed in isotropic turbulence, 
unless a dynamical mode reduction on helical modes with the same sign 
is imposed \cite{waleffe,biferale2012,biferale2013}. 
%%%%%%%   fig 1   %%%%%%
%%%%%%%%%%%%%%%%%%%%%%%%%%%%%%%%%%%%%%%%%%%%%%%%%%%%%%%%%%%%%%
\begin{figure*}[!htb]
\center
\hspace{-0.47cm}
\includegraphics[width=0.36\linewidth]{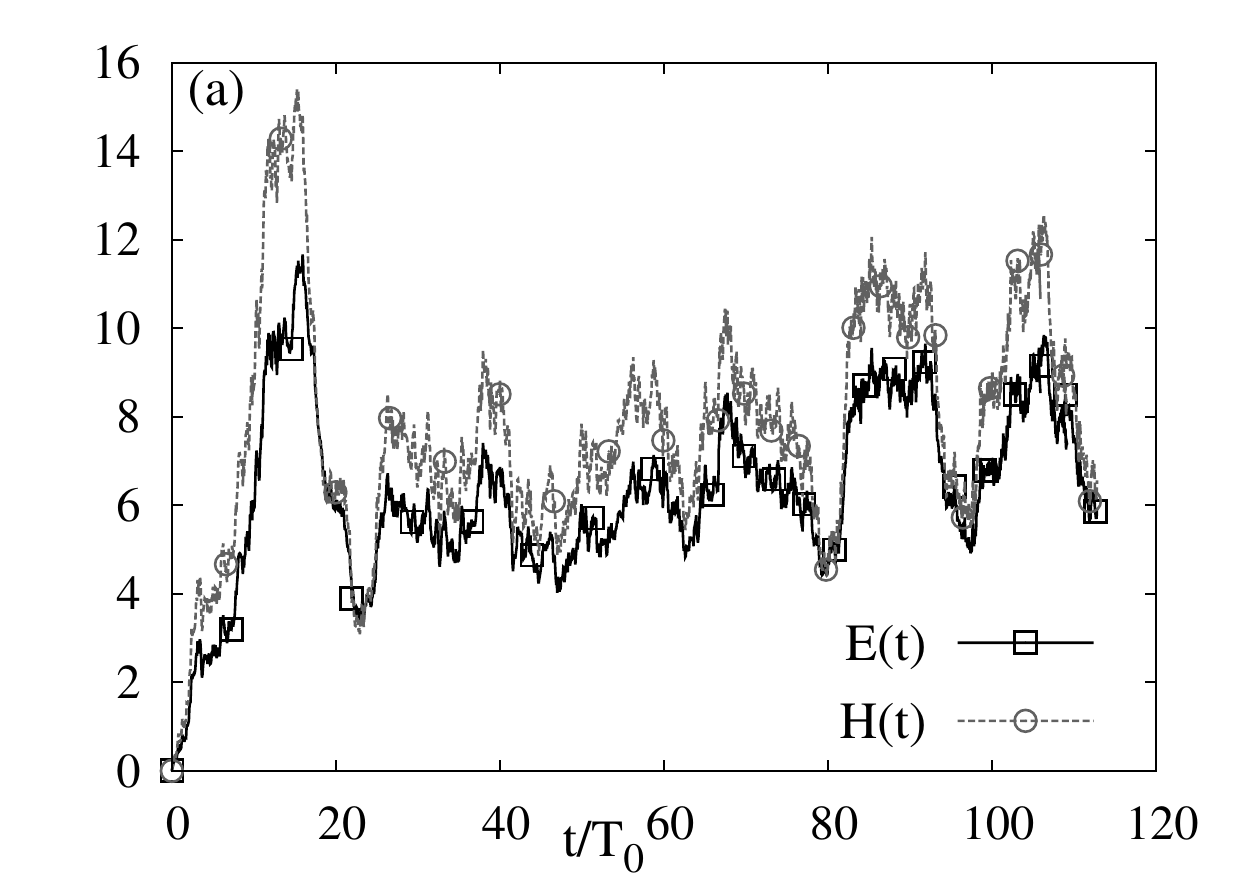}
\hspace{-0.65cm}
\includegraphics[width=0.36\linewidth]{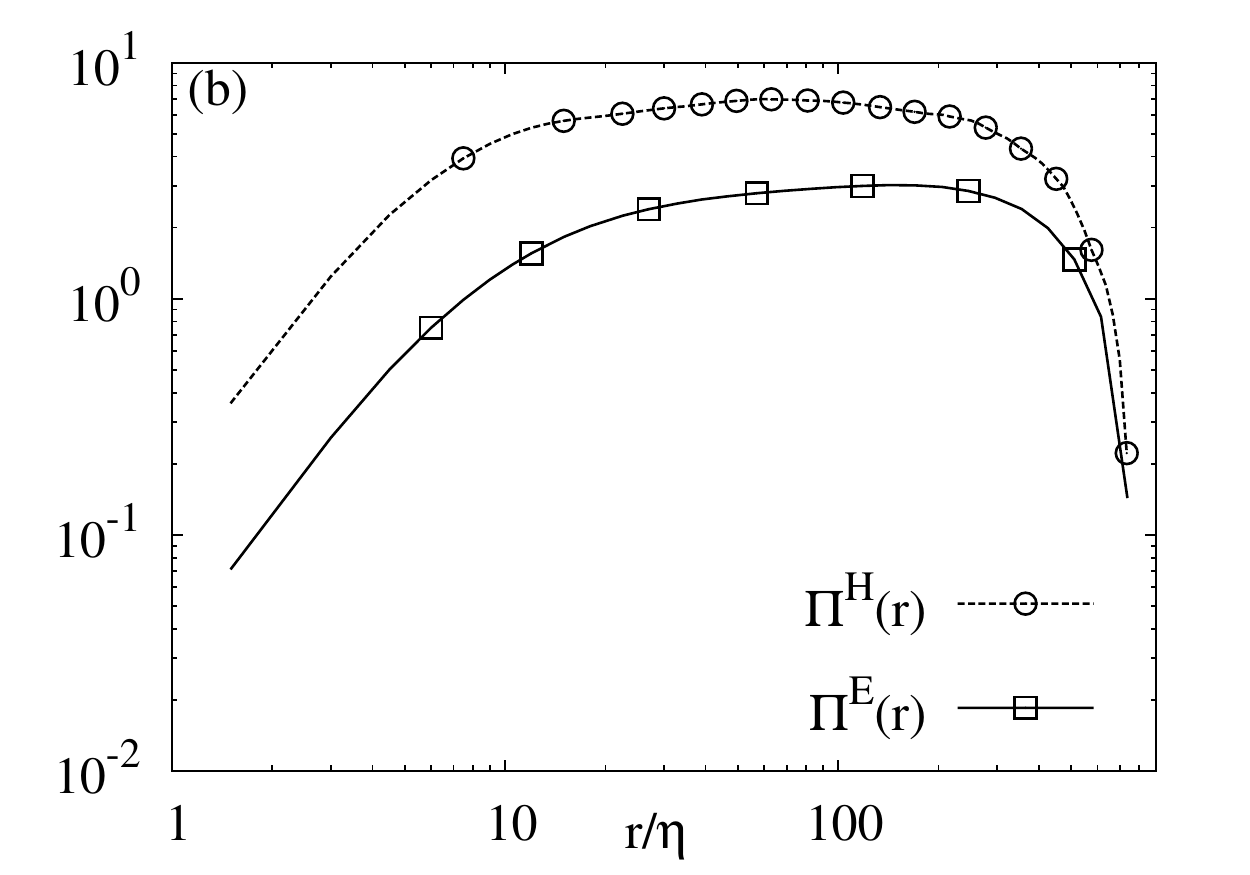}
\hspace{-0.65cm}
\includegraphics[width=0.36\linewidth]{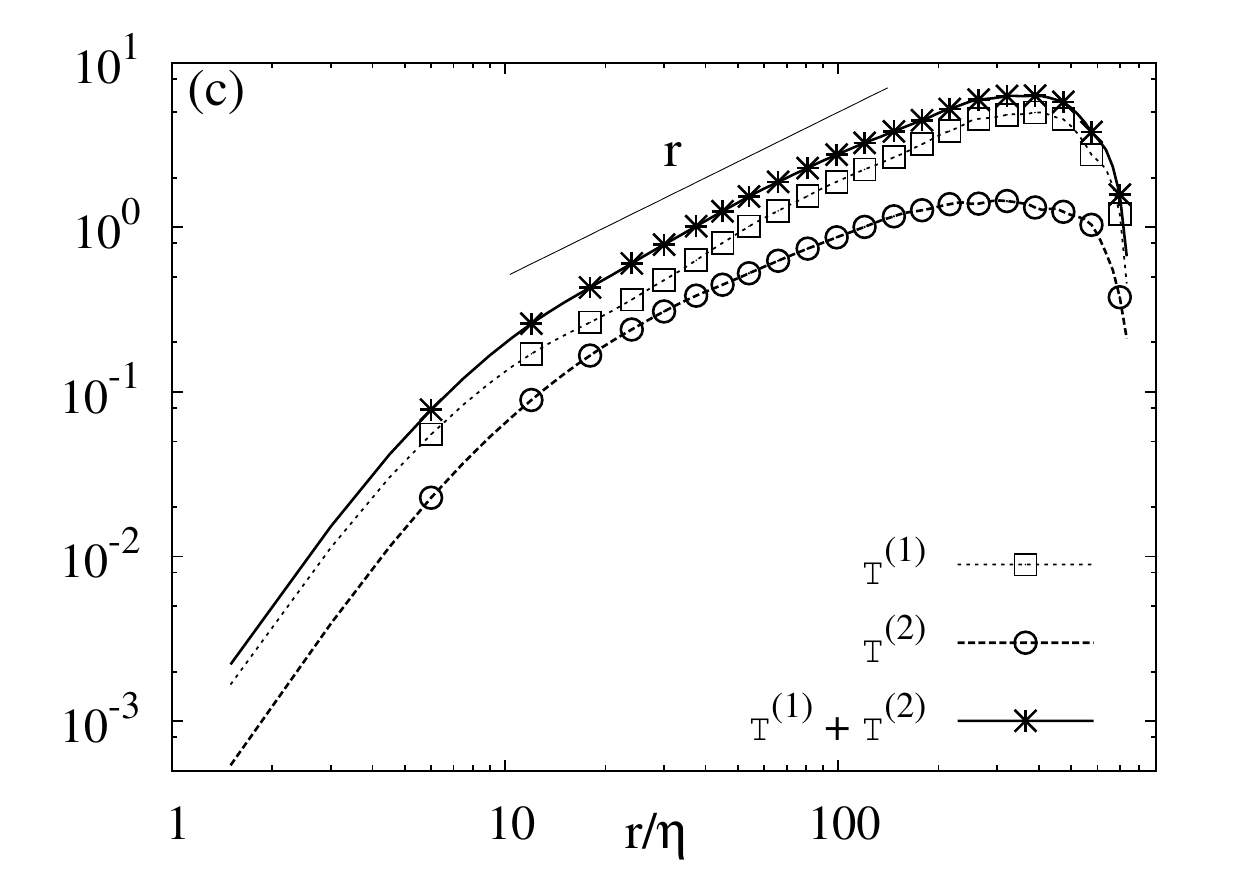}
\caption{(a) Temporal evolution of total energy and helicity from DNS
with injection of helicity (R1).  (b) Flux of energy $\Pi^E(r) = \langle
(\delta_r u)^3\rangle/r$ and flux of helicity $\Pi^H(r) = [{\cal
T}^{(1)}+{\cal T}^{(2)}]/r$, in real space, where ${\cal T}^{(1)}=\langle
\delta_r u (\delta_r \bm u \cdot \delta_r \bw)\rangle$ and ${\cal T}^{(2)}= -
0.5 \langle \delta_r \omega (\delta_r \bm u \cdot \delta_r \bm u)\rangle$ [see Eqs. 
 (\ref{eq:45})--(\ref{eq:45_second})]. (c) Scaling of the first term ${\cal
T}^{(1)}$, the second term ${\cal T}^{(2)}$, and their sum in 
(\ref{eq:45_second}).  The solid line is drawn with slope $1$ for comparison.} 
\label{fig:1}
\end{figure*}
%%%%%%%%%%%%%%%%%%%%%%%%%%%%%%%%%%%%%%%%%%%%%%%%%%%%%%%%%%%%%%
Besides the open issues concerning the spectral properties and higher-order 
statistics are even less studied and understood.  There are very few 
measurements of the mirror-antisymmetric components of structure 
functions. As a result, while a huge amount of work has been devoted to 
intermittency and anomalous scaling properties of the mirror-symmetric 
components, very little is known about the helical components 
\cite{chen2003prl, mininni_pouquet}.  In what follows we will  investigate 
further the statistical properties of helical turbulence concerning its 
spectral properties and beyond, assessing also the chiral components of 
high-order correlation functions. In particular, we will study the 
scaling properties of longitudinal structure functions based only on 
positive or negative helical modes: 
\begin{equation}
S^\pm_p(r) = \langle (\delta_ru^\pm)^p \rangle \, .
\end{equation}
In terms of the above decomposition, we can define energy- or 
helicity-like structure functions, i.e., combinations that are symmetric 
or antisymmetric for the exchange of  positive and negative helical 
projections: 
\begin{align}
&S^E_p(r) = \langle (\delta_ru^+)^p \rangle + \langle (\delta_ru^-)^p \rangle 
\sim r^{\zeta_p^E} \, , \label{eq:spsm} \\
&S^H_p(r) = \langle (\delta_ru^+)^p \rangle - \langle (\delta_ru^-)^p \rangle 
\sim r^{\zeta_p^H} \, . \label{eq:spsm_2}
\end{align} 
The advantage of working with the above definition is to avoid observables 
based on 
vorticity increments, which are strongly influenced by viscous contributions
and might not have a powerlaw scaling in the inertial range. 
In order to have a dimensional estimate for (\ref{eq:spsm})--(\ref{eq:spsm_2}) 
we start 
from the phenomenological predictions (\ref{eq:epm-chen}) considering  
the helical component to be subleading. Then, one might dimensionally write:
\begin{equation}
\label{eq:dim}
\delta_r u^\pm \sim \varepsilon^{1/3} r^{1/3} \pm h \varepsilon^{-2/3} r^{4/3} 
\, ,
\end{equation}
and therefore conclude that, at the lowest  order in $h$,
\begin{align}
&S^E_p(r) \sim \varepsilon^{\frac{p}{3}} r^{\frac{p}{3}} + o(h) \, , 
\label{eq:spsm2}\\
&S^H_p(r) \sim \varepsilon^{\frac{p}{3}-1} h r^{\frac{p}{3}+1} + O(h^2) \, , 
\label{eq:spsm2_2}
\end{align} 
where the second relation is obtained taking into account that the 
leading terms proportional to $\varepsilon^{p/3}$ cancel out. \\ Another 
possible way to highlight the scaling properties of the helical  
component of the {\it scale-by-scale} velocity statistics  is to look 
directly at the local helicity increments:
\begin{equation}
\label{eq:hsf}
{\cal H}_p(r) = \langle \sign(\delta_r u_i \delta_r \omega_i) |\delta_r u_i 
\delta_r \omega_i|^p\rangle,
% \sim r^{\zeta_p^{\cal H}},
\end{equation}
where we have introduced the $\sign$ function in 
order to have a chiral observable for all orders of the 
moment $p$ \footnote{Notice that in Ref. \cite{mininni_pouquet} these structure 
functions are defined without the $\sign$ and hence are helicity-sensitive only 
for odd moments.}\cite{ditlevsen2000}.
The mean value of the {\it sign}
of local helicity, which gives a direct measure of the relative importance of 
chiral-fluctuations with respect to the non-chiral background, 
is known as the {\it cancellation exponent} \cite{cancellation}, and  
can be estimated dimensionally to be:
\begin{align}
{\cal H}_0(r) = \left< \frac{ \delta_r u_i \delta_r \omega_i}{ |\delta_r u_i 
\delta_r \omega_i|} 
\right> \sim \frac{h \varepsilon^{-1/3} r^{2/3}}{\varepsilon^{2/3} r^{1/3} 
\eta^{-2/3}} \sim h \varepsilon^{-1} r^{1/3}\eta^{2/3}  \,,
\label{eq:hsf0}
\end{align}
where we write the numerator in terms of its dominant helical 
contribution and the denominator as the mirror-symmetric term with 
$|\delta_r u| \sim \varepsilon^{1/3} r^{1/3}$ and $|\delta_r \omega| \sim 
\varepsilon^{1/3} \eta^{-2/3}$. Here $\eta$ is the Kolmogorov 
length-scale, where the vorticity increment is expected to be maximal. 
As a result, we should have for (\ref{eq:hsf}) the scaling property: 

\begin{equation}
{\cal H}_p(r) \sim  h 
\eta^{\frac{2-2p}{3}}\varepsilon^{\frac{2p-3}{3}}  r^{\frac{p+1}{3}} \, . 
\label{eq:prediction_h_cal}
\end{equation}

%%%%%%%%%%%%%%%%%%%%% table 1 %%%%%%%%%%%%%%%%%%%%%%%%%%%%%%%%
%$\lambda = u_{\rm rms}/\sqrt{\langle\left[\partial_x u(\bx)\right]^2\rangle}$
\begin{table}[b]
  \caption{Details of the simulations. $N$: number of collocation points along
each axis; $L$: size of the periodic box; $\nu$: kinematic viscosity; $k_f$:
range of forced wavenumbers; $u_{\rm rms}$: rms velocity; $Re_\lambda= 
u_{\rm rms} \lambda/\nu$: Taylor-microscale Reynolds number,
 where $\lambda = \frac{2\pi}{L}\sqrt{\frac{\langle u^2(\bx) 
 \rangle}{\langle\left[\partial_x u(\bx)\right]^2\rangle}}$ is the Taylor 
 microscale; $\varepsilon$: mean
energy dissipation rate; $\eta$: Kolmogorov length-scale; $T_0$:
large-eddy-turnover time.}
  \label{table1}
\begin{tabular*}{\linewidth}{@{\extracolsep{\fill} } c  c  c  c  c  c  c  c  c  
c}
    \toprule
   Run&  $N$   & $L$ & $\nu$   & $k_f$   & $u_{\rm rms}$ & $Re_\lambda$ 
   &$\varepsilon$ & $\eta$  & $T_0$ \\ \colrule
   R1 & $1024$ & $2\pi$ & $0.001$ & $[1,2]$ & $3.4$      &  $320$       &    
   $3.2$     & $0.004$ & $0.3$ \\
    \botrule
  \end{tabular*}
\end{table}

%Here I used urms = sqrt(2E); T_0 = L0/urms; L0 is 6th column in the 
%checkfluid.dat

%%%%%%%%%%%%%%%%%%%%%%%%%%%%%%%%%%%%%%%%%%%%%%%%%%%%%%%%%%%%%%

\section{Numerical Simulations of Navier-Stokes equations \label{section:NS}} 
%%%%%%%%%%%%%%%%%%%%%%%%%%%%%%%%%%%%%%%%%%%%%%%%%%%%%%%%%%%%%%
%%%%%%%   fig 2   %%%%%%
%%%%%%%%%%%%%%%%%%%%%%%%%%%%%%%%%%%%%%%%%%%%%%%%%%%%%%%%%%%%%%
\begin{figure}[!t]
\center
\includegraphics[width=0.63\linewidth]{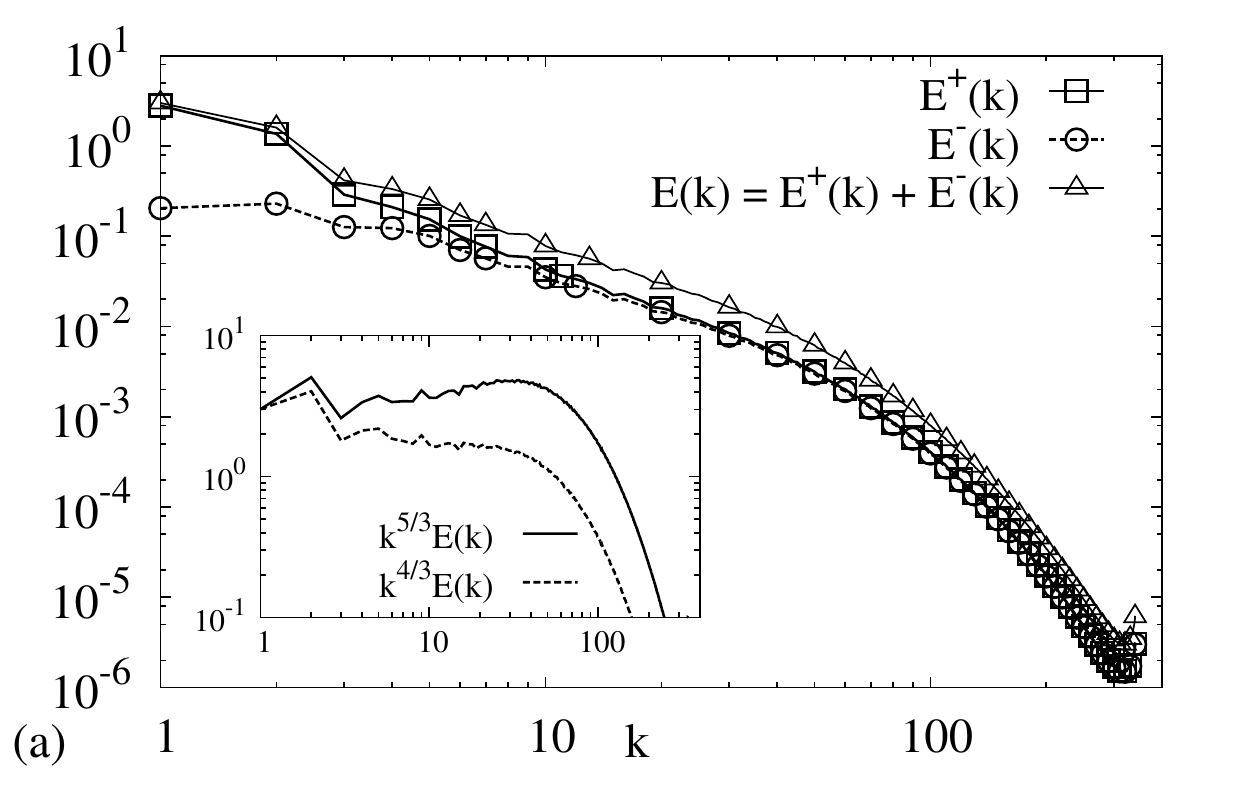}
\includegraphics[width=0.63\linewidth]{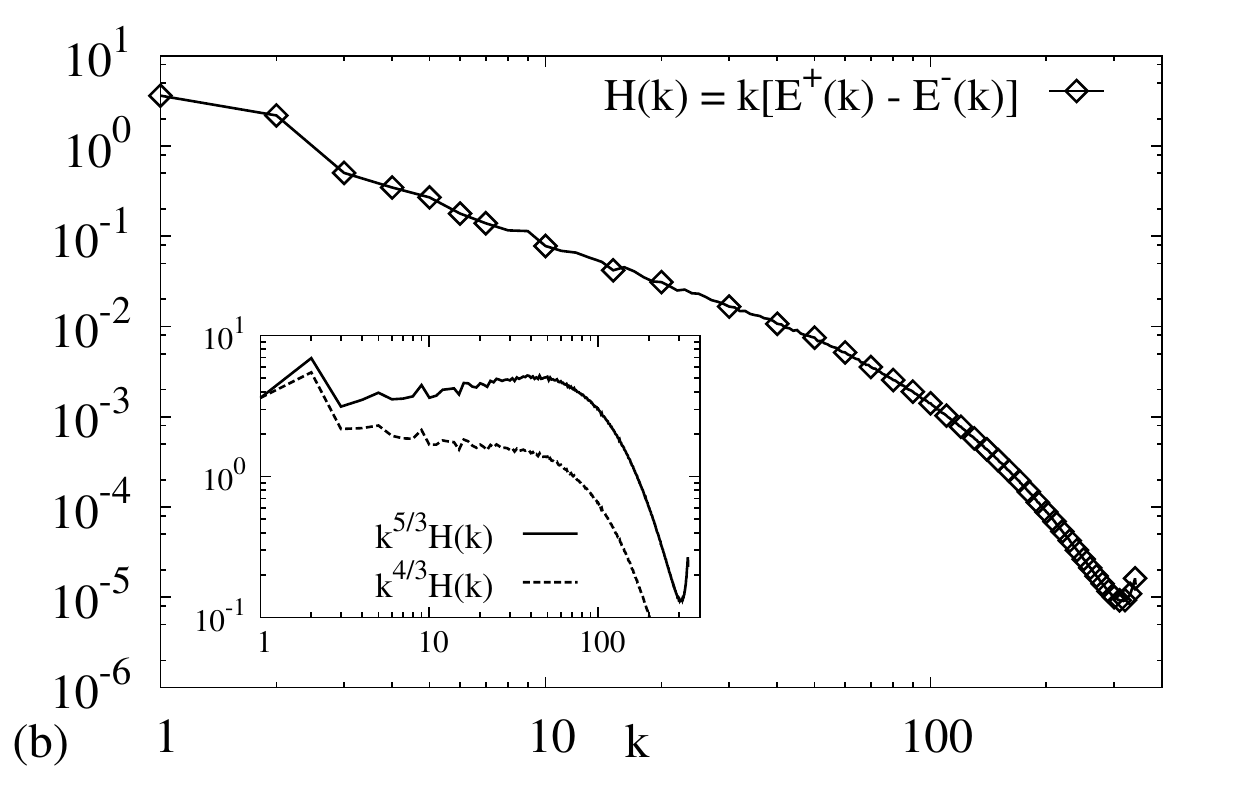}
\caption{(a) Log-log plots of energy spectrum $E(k)$ and its helical components 
$E^\pm(k)$. 
Inset: Compensated energy spectra with predictions (\ref{eq:epm-chen}) and 
(\ref{eq:epm-kurien}).
 (b) Log-log plots of helicity spectrum $H(k)$. Inset: Compensated helicity
 spectra with predictions (\ref{eq:epm-chen}) and (\ref{eq:epm-kurien}).}
\label{fig:2}
\end{figure}
%%%%%%%%%%%%%%%%%%%%%%%%%%%%%%%%%%%%%%%%%%%%%%%%%%%%%%%%%%%%%%
%%%%%%%   fig 3   %%%%%%
%%%%%%%%%%%%%%%%%%%%%%%%%%%%%%%%%%%%%%%%%%%%%%%%%%%%%%%%%%%%%%
\begin{figure}[!t] 
\center 
\includegraphics[width=0.63\linewidth]{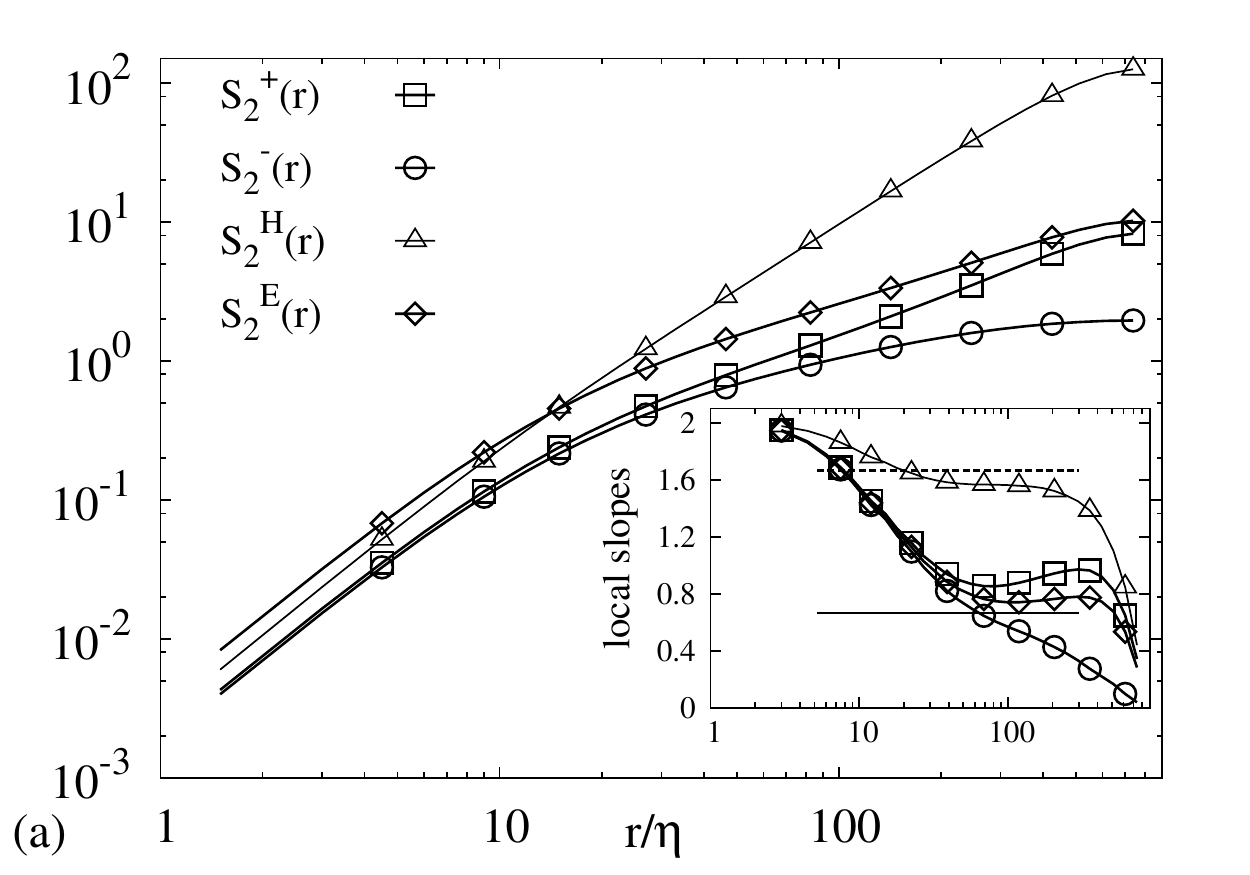}\\ \vspace{-0.2cm}
\includegraphics[width=0.63\linewidth]{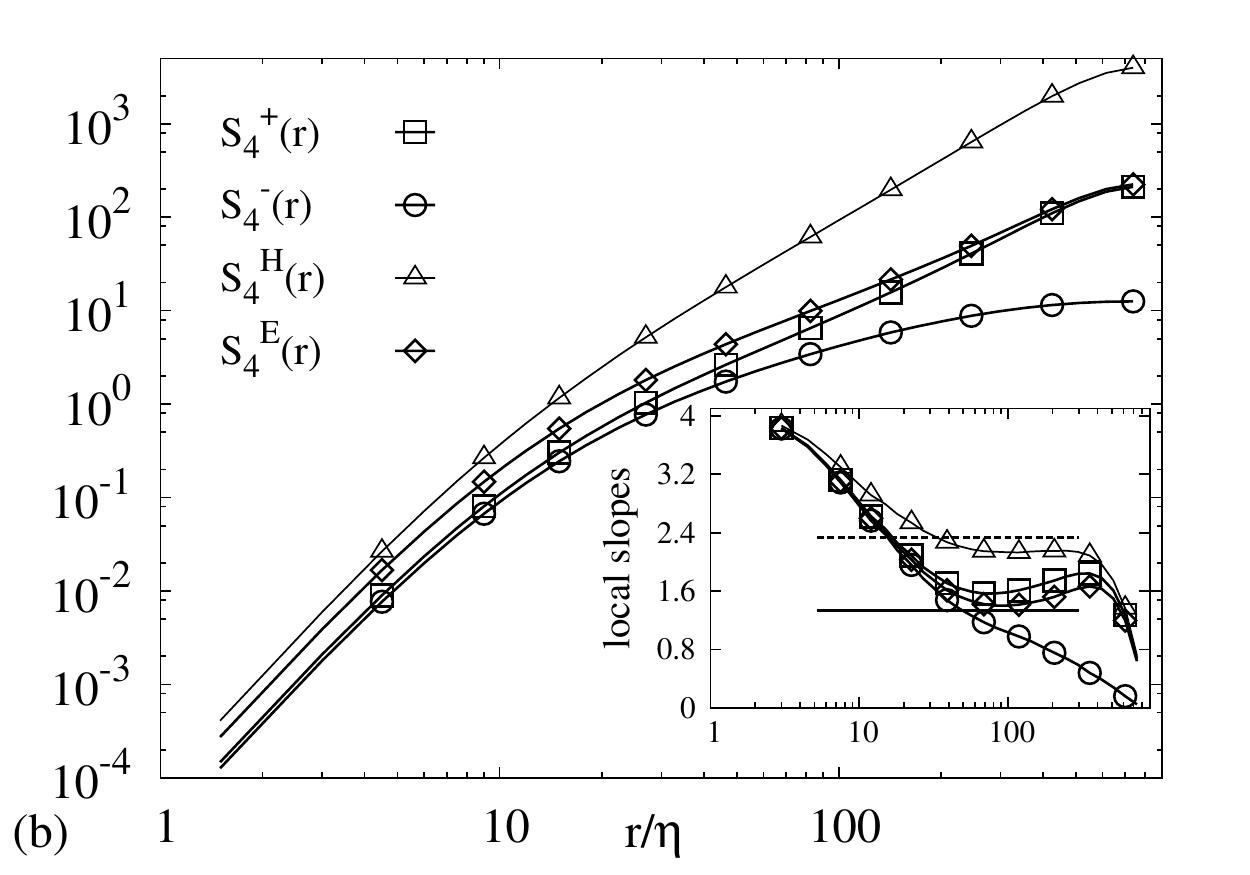}\\ \vspace{-0.2cm}
\includegraphics[width=0.63\linewidth]{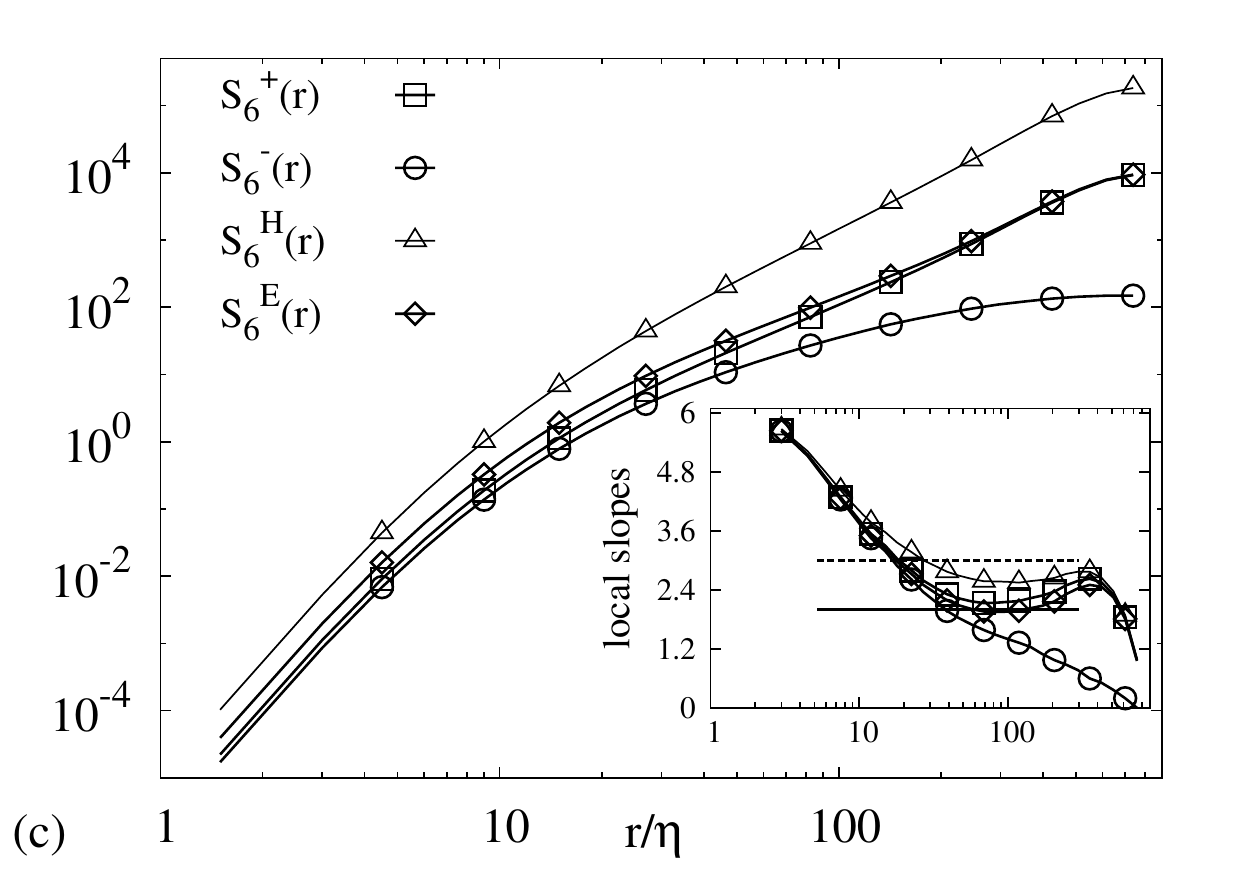}
\caption{Log-log plots of structure functions in real space for (a) second, (b)
fourth, and (c) sixth order based on positive or negative, $S_p^\pm(r)$,  helical
modes and their combinations, $S_p^{E,H}(r)$. $S_p^H(r)$ are multiplied with a
scalar factor for better representation. Inset: Local slopes of the curves
showing $\zeta^{E,H}_p(r)$  and $\zeta^{\pm}_p(r)$ [see (\ref{eq:zeta_eh})].
Dimensional predictions (solid lines for $p/3$ and dashed lines for
$\frac{p}{3}+1$) are also shown for comparison, where $p$ is the order of the
structure functions.}
\label{fig:3} 
\end{figure}
%%%%%%%   fig 4   %%%%%%
%%%%%%%%%%%%%%%%%%%%%%%%%%%%%%%%%%%%%%%%%%%%%%%%%%%%%%%%%%%%%%
\begin{figure}[!htb]
\includegraphics[width=0.63\linewidth]{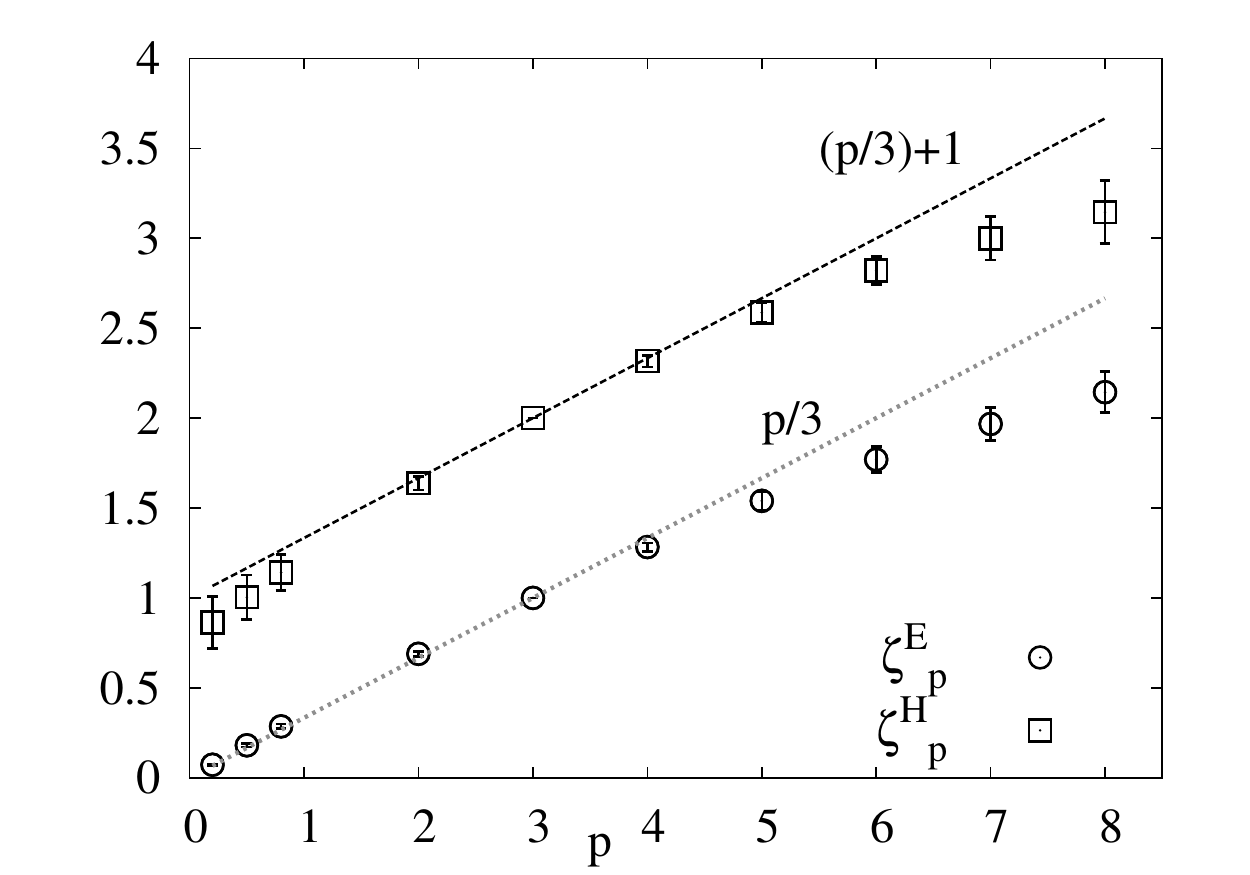}
\caption{Scaling exponents $\zeta_p^{E,H}$ of the chiral-symmetric and
chiral-antisymmetric structure functions obtained using ESS.  The lines with 
slopes
$\frac{p}{3}$ and $\frac{p}{3}+1$ correspond to the dimensional predictions for
$\zeta_p^{E}$ and $\zeta_p^{H}$, respectively.  The errorbars show the
variation of the exponents within the inertial range.} 
  \label{fig:4}
\end{figure}
%%%%%%%%%%%%%%%%%%%%%%%%%%%%%%%%%%%%%%%%%%%%%%%%%%%%%%%%%%%%%%

%%%%%%%   fig 5   %%%%%%
%%%%%%%%%%%%%%%%%%%%%%%%%%%%%%%%%%%%%%%%%%%%%%%%%%%%%%%%%%%%%%
\begin{figure*}[!htb] 
\center 
\includegraphics[width=0.49\linewidth]{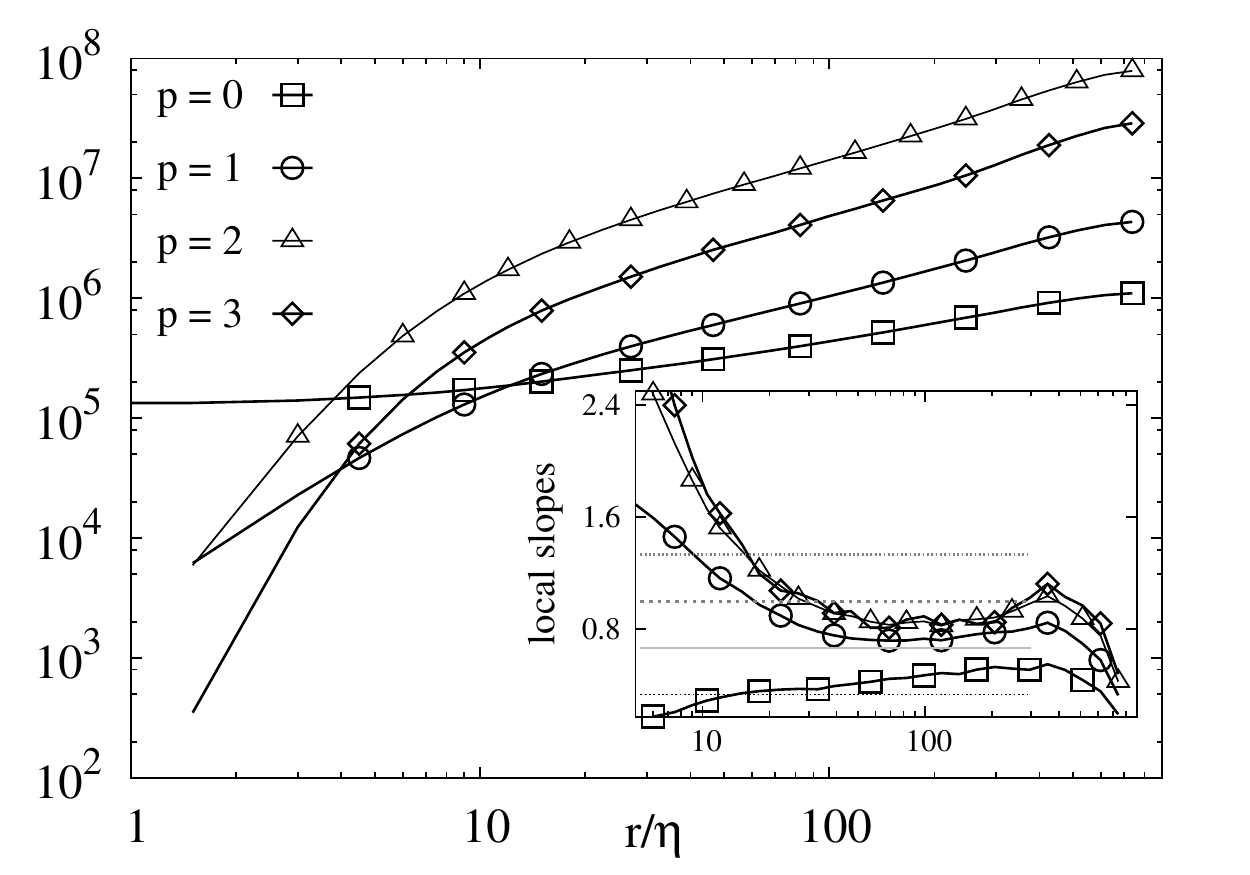}
\includegraphics[width=0.49\linewidth]{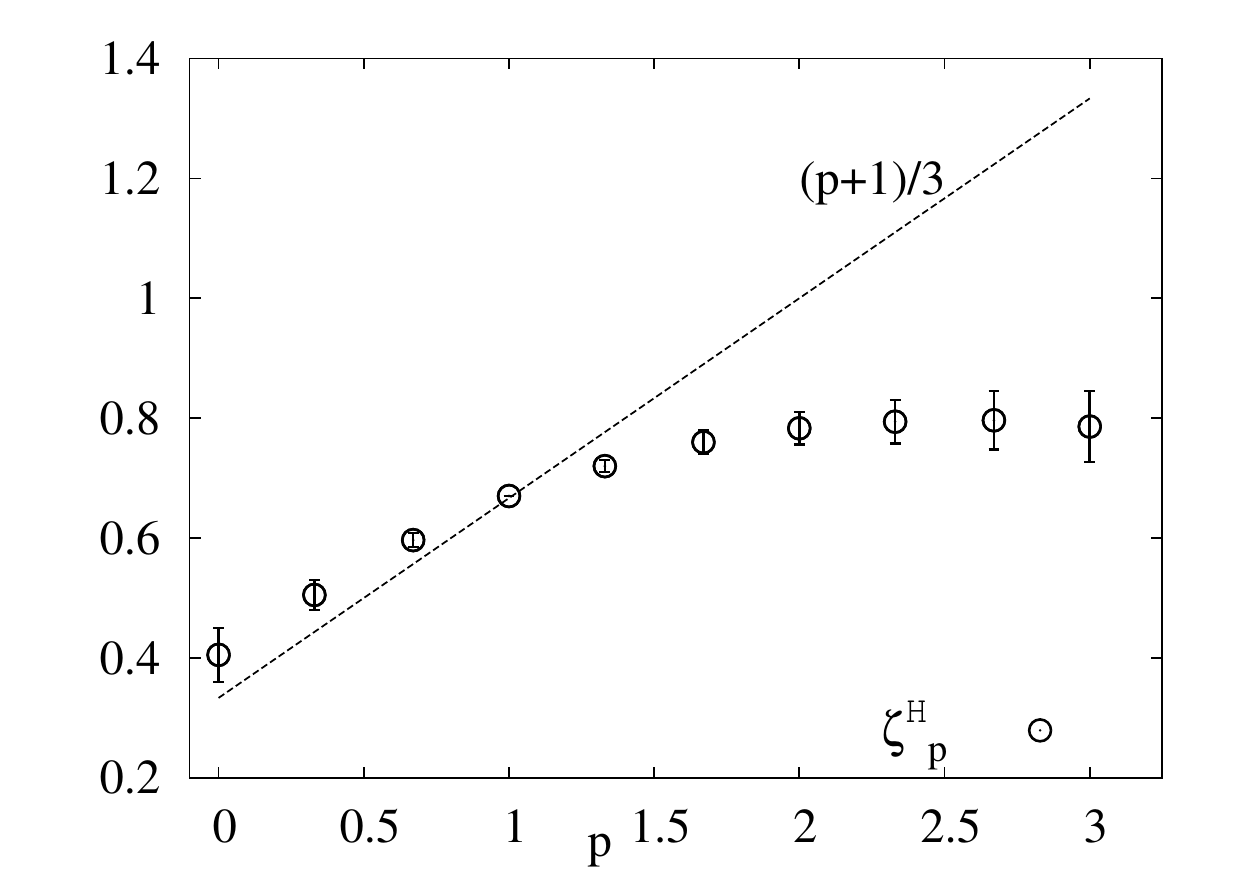}
  \caption{(Left): Log-log plots of helicity structure functions ${\cal
H}_p(r)$, defined in (\ref{eq:hsf}), for $p=0$, $1$, $2$ and $3$. Inset: local
slopes of the same curves. Horizontal lines indicate the values $\frac{1}{3}$,
$\frac{2}{3}$, $1$, and $\frac{4}{3}$ from bottom to top.  (Right): Scaling
exponents $\zeta_p^{\cal H}$ of ${\cal H}_p(r)$ obtained using ESS. The line 
with slope $\frac{p+1}{3}$ corresponds to
the dimensional prediction.  The errorbars show the variation of the exponents
within the inertial range.} 
\label{fig:5} 
\end{figure*}
%%%%%%%%%%%%%%%%%%%%%%%%%%%%%%%%%%%%%%%%%%%%%%%%%%%%%%%%%%%%%%

We have performed a series of DNSs of the NSE (\ref{eq:ns+++}) with a
fully-dealiased, pseudospectral code at a resolution of $1024^3$ collocation
points on a triply periodic cubic domain of size $L=2\pi$.  The flow is
sustained by a random Gaussian forcing with \[\langle f_i(\bk,t) f_j(\bq,t')
\rangle = F(k) \delta(\bk-\bq) \delta(t-t') Q_{i,j}(\bk),\] where $Q_{ij}(\bk)$
is a projector imposing  incompressibility and $F(k)$ has support only for $
k_f \in [k_{\rm min}, k_{\rm max}]$. We carried out a DNS, namely,  R1,
where we inject maximal helicity by forcing only the positive helical modes of
the velocity (see Table~\ref{table1}).  

In Fig.~\ref{fig:1}(a) we show the evolution of the total energy $E$ and 
helicity
$H$, and of their fluxes in real space [Fig. \ref{fig:1}(b)]. It is clear that the system 
is in a
stationary state with a dual forward cascade of energy and helicity. 
Figure \ref{fig:1}(c) shows that the two contributions entering in
(\ref{eq:45_second}) have different properties in the inertial range
and that only their sum shows a good linear behavior as predicted by
the constant-helicity-flux solution. This is not surprising; we must expect
that in the presence of two transferred quantities only particular combinations of
correlation functions might have an exact scaling behavior, while any general
combination of fields might be affected by leading and subleading
contributions. 

In Fig.~\ref{fig:2} we show the energy and helicity spectra and their 
positive and negative helical components $E^\pm(k)$. We observe that 
the predictions (\ref{eq:chen1})--(\ref{eq:chen2}) give a better 
compensation at least for not too high wavenumbers where a dissipative
bottleneck is known to affect the local scaling properties. At
those wavenumbers, the relative helicity $H(k)/kE(k)$ is already very small
and it is unlikely that the bottleneck is due to some helical effects as
proposed by (\ref{eq:epm-kurien}). 
Concerning real-space quantities, in Fig.~\ref{fig:3} we show the scaling of  
structure functions of positive and negative helical components of velocity 
together with their combinations for the second, fourth and sixth order.
To verify the dimensional scaling predictions 
(\ref{eq:spsm2})--(\ref{eq:spsm2_2})
we calculate the local slopes of $S_p^{E,H}(r)$ and $S_p^\pm(r)$: 
\begin{align}
\label{eq:zeta_eh}
 \zeta_p^{E,H}(r) = \frac{d\, \log S_p^{E,H}(r)}{d\, \log r },\;\; 
 \zeta_p^{\pm}(r) = \frac{d\, \log S_p^{\pm}(r)}{d\, \log r },
\end{align}
as shown in the insets of Fig.~\ref{fig:3}.  We then used ESS \cite{ess_benzi} to obtain  a better fit of the
relative scaling exponents in the inertial range:
$\zeta_p^{E,H}/\zeta_3^{E,H}$.  In Fig.~\ref{fig:4} we compare the scaling
exponents $\zeta_p^{E,H}$ and their dimensional predictions
(\ref{eq:spsm2})--(\ref{eq:spsm2_2}). To derive the absolute value of the
scaling exponents out of the ESS scaling we have assumed, $\zeta_3^E=1$ and
$\zeta_3^H=2$ in agreement with the exact scaling properties
(\ref{eq:45})--(\ref{eq:45_second}). From this figure we can see that the
dimensional prediction is well verified, except for the presence of a small
anomalous correction for high-order moments. 

In Fig.~\ref{fig:5} we show
the scaling behavior of ${\cal H}_p(r)$ for values of $p$ from $0$ to $3$
compared with the dimensional prediction (\ref{eq:prediction_h_cal}). This works
well up to $p \sim 1.5$, while for $p \ge 2 $ nontrivial anomalous corrections
appear.
 
In summary,
the scaling exponents for the two sets of helicity-sensitive structure
functions $S_p^H(r)$  agree well  with the dimensional estimate
except for a small anomalous correction which is of the same order of the one
observed for the mirror-symmetric terms. One might argue that the two set of
anomalous exponents should be correlated, being connected to the dependency on
the energy dissipation on the right-hand side of
(\ref{eq:spsm2})--(\ref{eq:spsm2_2}).
It is important to notice that helicity is not positive definite, and
its dissipation can be split in two different channels, one for positive and
one for negative helical components. The theoretical dependency on the Reynolds
number of the two processes and of the total helicity dissipation is discussed
in Refs. \cite{chen2003joint,ditlevsen2000,ditlevsen_book}. Further studies of
changing Reynolds number would be needed to clarify the existence of a
dissipative anomaly for the helicity cascade and the dependency of the whole
statistics on the turbulence intensity.  The multiscale nature of the
correlation involving vorticity and velocity in ${\cal H}_p(r) $ might be
particularly sensitive to fluctuations of the dissipative physics, hence
explaining the large intermittent correction shown in Fig. \ref{fig:5}.

In order to investigate further the statistics of the helicity transfer
 we present in the next section a study of helical shell
models, where it is possible to considerably increase the Reynolds number. 
\section{Helical Shell Models}
%%%%%%%   fig 6   %%%%%%
%%%%%%%%%%%%%%%%%%%%%%%%%%%%%%%%%%%%%%%%%%%%%%%%%%%%%%%%%%%%%%
\begin{figure}[t]
\center
\includegraphics[width=0.63\linewidth]{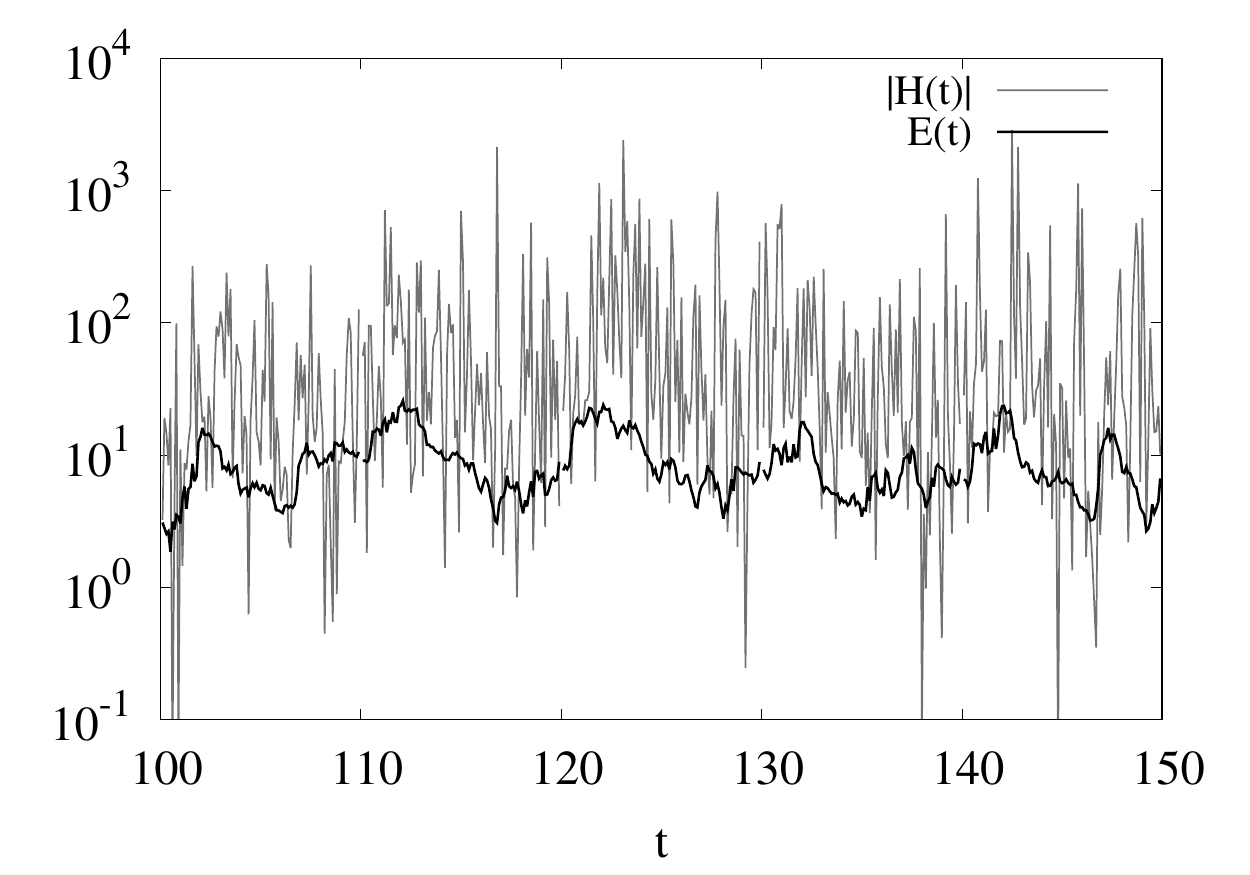}
\caption{Time evolution of total energy (\ref{eq:energy_def}) and total
helicity (\ref{eq:helicity_def}) in a typical run of the shell model 
simulation.}
\label{fig:6} 
\end{figure} 
%%%%%%%%%%%%%%%%%%%%%%%%%%%%%%%%%%%%%%%%%%%%%%%%%%%%%%%%%%%%%%
To check the robustness of the previous findings, we studied the same helical
structure functions in a family of helical shell models
\cite{biferale_helical_1996}. Shell models have been useful to study cascade
processes and scaling behaviors  in
turbulent flows since they allow us to achieve very high Reynolds numbers in
numerical simulations \cite{jensen_dynamical_systems_book,lvov_improved, 
ditlevsen1997, biferale2003shell, ditlevsen_book, ssray_njp,stepanov_prl}.

Shell models are based on  a simplified dynamical evolution of the 
energy and helicity transfer by keeping only one (or a few) modes for each 
spherical 
shell in Fourier space. They represent a drastic non exact reduction of 
the degrees-of-freedom of the NSE. The original 
idea is to describe the evolution of a single complex variable $u_n$, 
representing all the modes in a shell of wavenumbers $ k \in 
[k_n,k_{n+1}]$, with $k_n$ equispaced in logarithmic scale, $k_n = 2^n 
k_0$.  The first step to have a realistic helical structure was done in Ref. 
\cite{biferale_helical_1996}, where two complex variables $u_n^+$ 
and $u_n^-$ carrying positive or negative helicity were introduced for every 
wavenumber.
This lead to four independent classes of helical 
shell models, mimicking exactly the four classes of helical 
interactions of the original NSE \cite{waleffe}. 
Other models based on similar decompositions have also been proposed 
\cite{plunian_stepanov_frick_review, rathmann_ditlevsen_2016, 
gledzer2015inverse}. 
Here we follow the structure given in \cite{biferale_helical_1996}, 
where the four possible models have the general form:
\begin{equation}
\begin{aligned}
\label{eq:sabra_helical_standard}
\dot{u}_n^+ = & \, i (a k_{n+1} u_{n+2}^{s_1} u_{n+1}^{s_2*} + b k_{n} 
u_{n+1}^{s_3} u_{n-1}^{s_4*} \\
& + c k_{n-1} u_{n-1}^{s_5} u_{n-2}^{s_6}) + f_n^+ - \nu k_n^2  u_n^+  \, ,\\
\dot{u}_n^- = & \, i (a k_{n+1} u_{n+2}^{-s_1} u_{n+1}^{-s_2*} + b k_{n} 
u_{n+1}^{-s_3} u_{n-1}^{-s_4*} \\
& + c k_{n-1} u_{n-1}^{-s_5} u_{n-2}^{-s_6}) + f_n^-  - \nu k_n^2   u_n^-   \, .
\end{aligned}
\end{equation}
The helicity indices $s_i = \pm$ are reported in Table 
\ref{tab:helicity_indices} and the coefficients $a$, $b$, and $c$ can be found 
in 
Table \ref{tab:model_coefficients}.

%%%%%%%%%%%%%%%%%%%%%%%%%%%%%%%%%%%%%%%%%%%%%%%%%%%%%%%%%%%%%%
%%%%%%   fig 7   %%%%%%
%%%%%%%%%%%%%%%%%%%%%%%%%%%%%%%%%%%%%%%%%%%%%%%%%%%%%%%%%%%%%%
\begin{figure*}[ht]
\includegraphics[width=.49\linewidth]{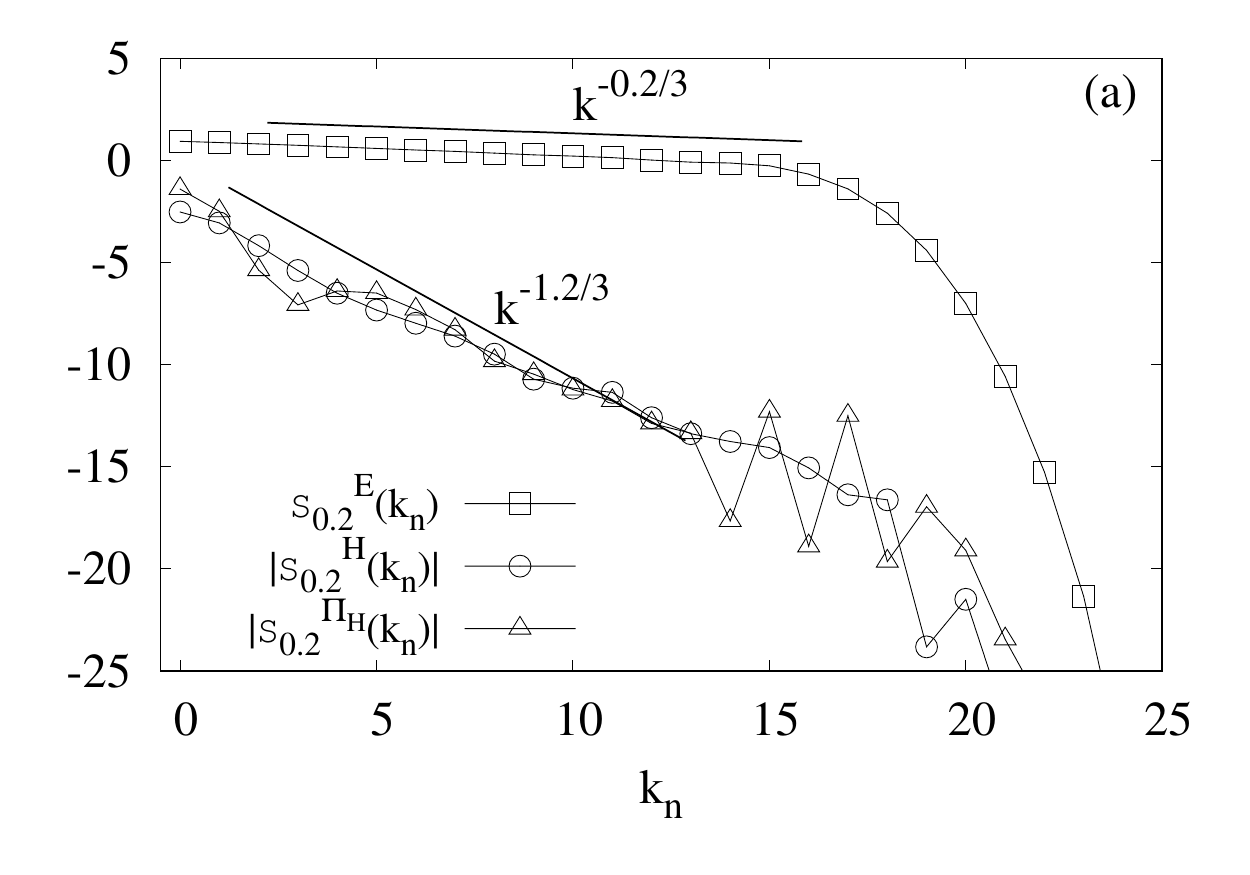}
\includegraphics[width=.49\linewidth]{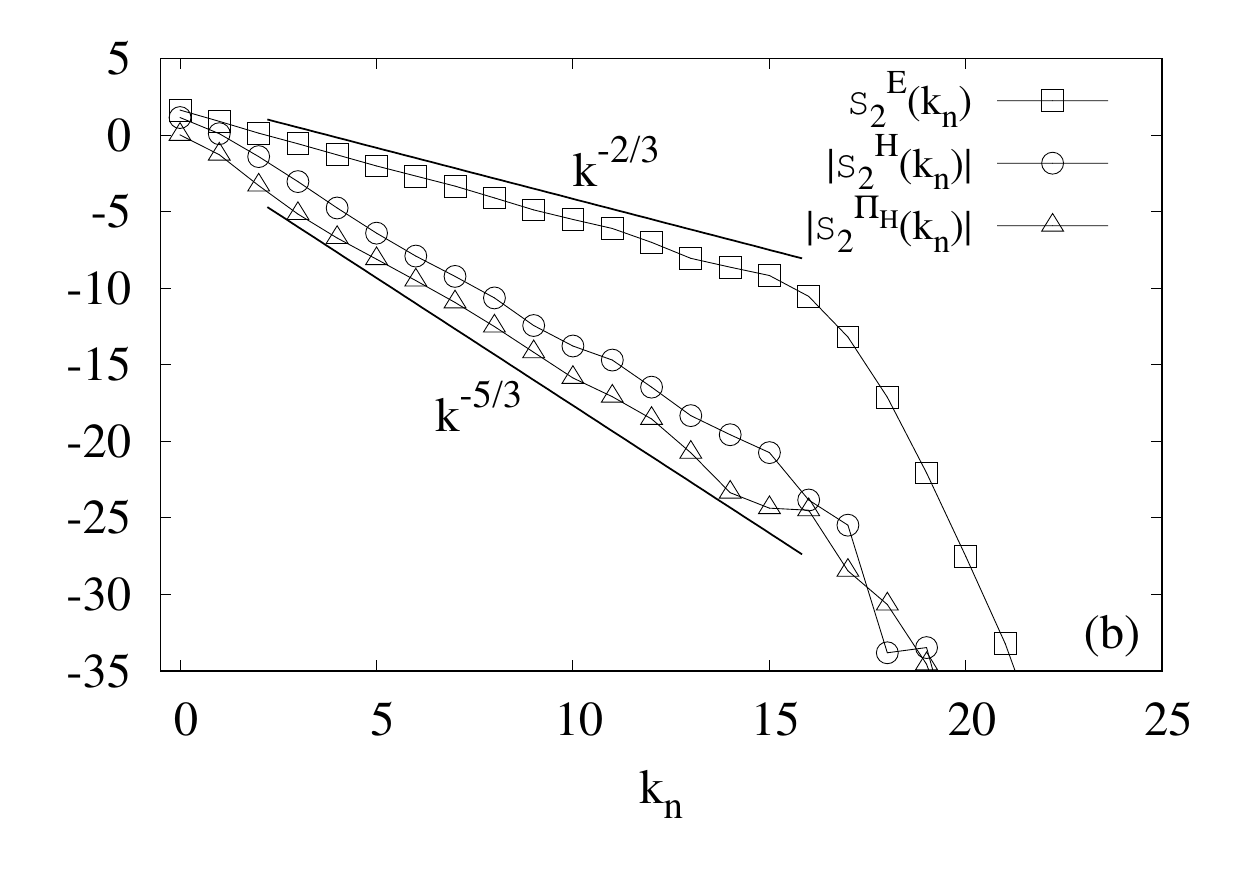}\\
\vspace{-0.5cm}
\includegraphics[width=.49\linewidth]{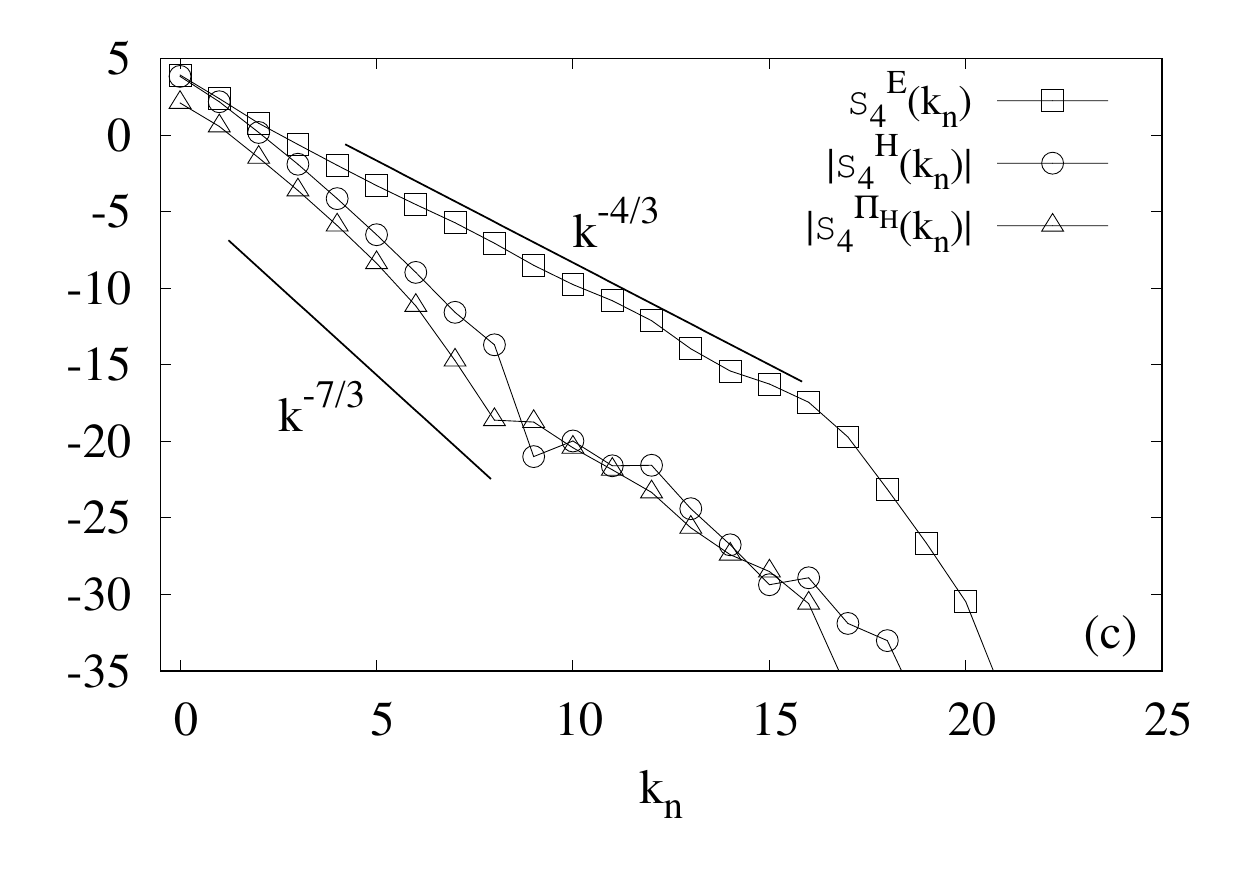}
\includegraphics[width=.49\linewidth]{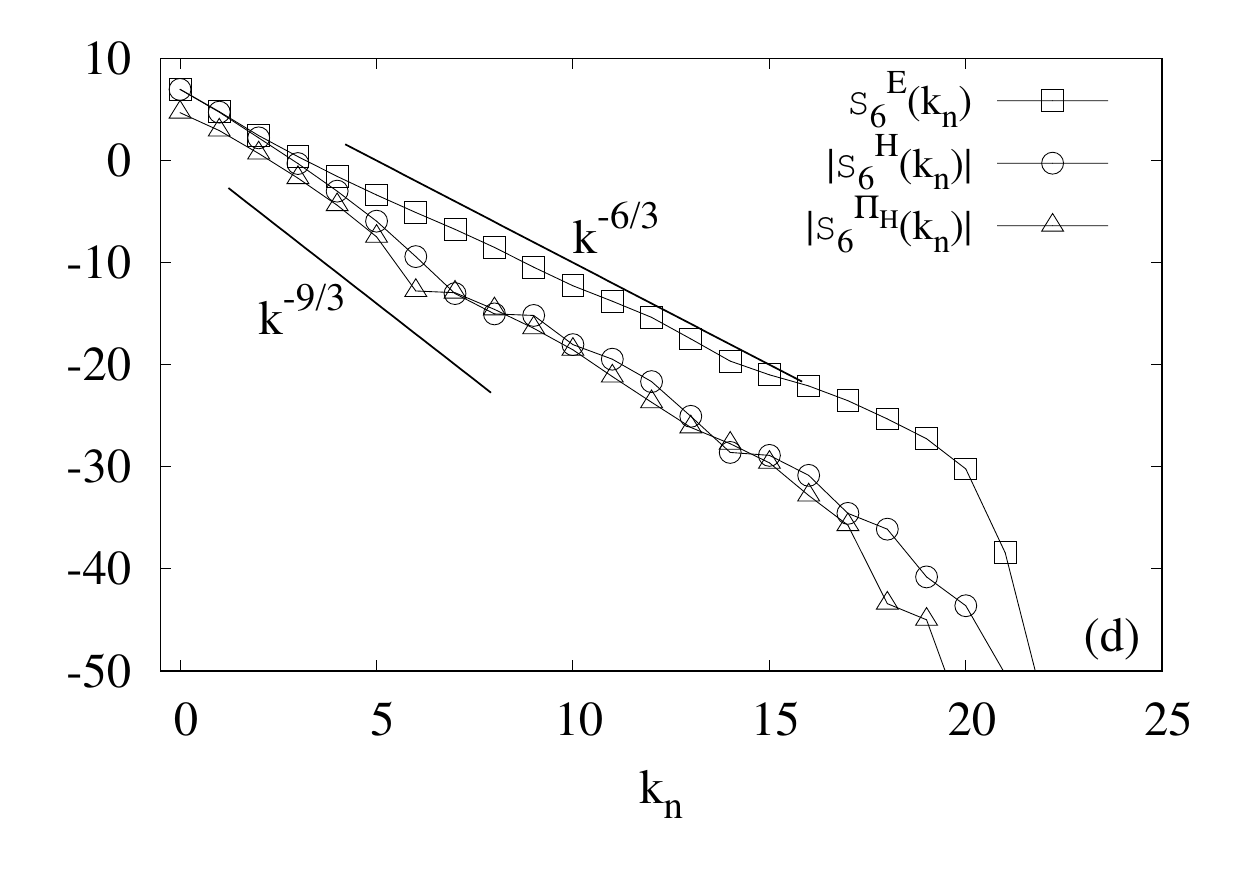}
\caption{Log-log (base 2) plots of the symmetric and antisymmetric 
(\ref{eq:S_H})-(\ref{eq:S_PI_H}) structure functions of different orders
$p=0.2$ (a), $p=2$ (b), $p=4$ (c), $p=6$ (d). Straight lines correspond to the
predictions
(\ref{eq:shell_structure_prediction_E})-(\ref{eq:shell_structure_prediction_H}).}
\label{fig:7}
\end{figure*} 
%%%%%%%%%%%%%%%
%%%%%%%   fig 8   %%%%%%
%%%%%%%%%%%%%%%%%%%%%%%%%%%%%%%%%%%%%%%%%%%%%%%%%%%%%%%%%%%%%%
\begin{figure}[htb]
\includegraphics[width=0.63\linewidth]{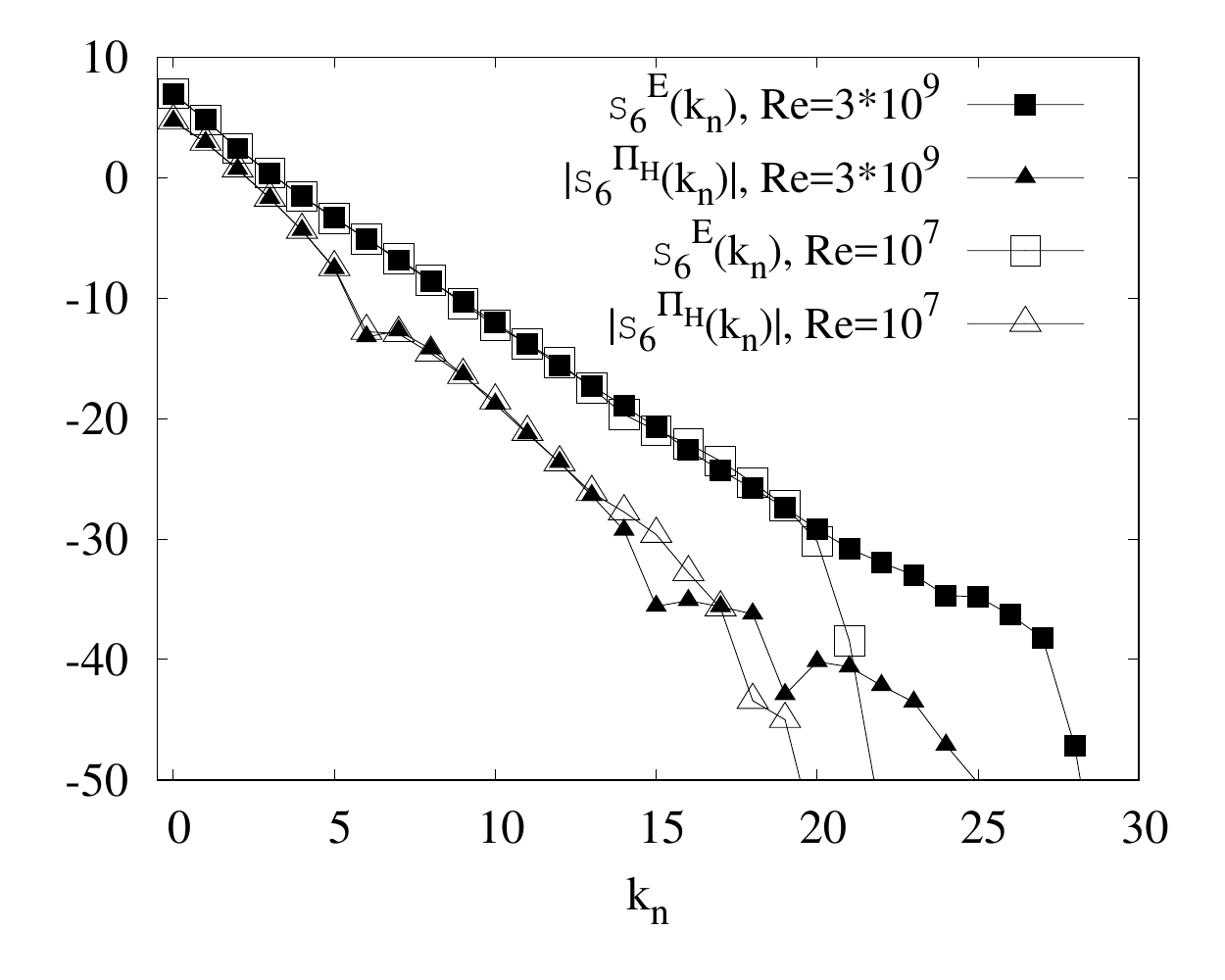}
\caption{Log-log (base 2) plot of the sixth order structure functions for two 
simulations at different Reynolds numbers: $Re \sim 10^7$ and $Re \sim 
3 \cdot 10^9$.}
\label{fig:8} 
\end{figure} 
%%%%%%%%%%%%%%%%%%%%%%%%%%%%%%%%%%%%%%%%%%%%%%%%%%%%%%%%%%%%%%
%%%%%%%   fig 9   %%%%%%
%%%%%%%%%%%%%%%%%%%%%%%%%%%%%%%%%%%%%%%%%%%%%%%%%%%%%%%%%%%%%%
\begin{figure}[htb]
\includegraphics[width=0.63\linewidth]{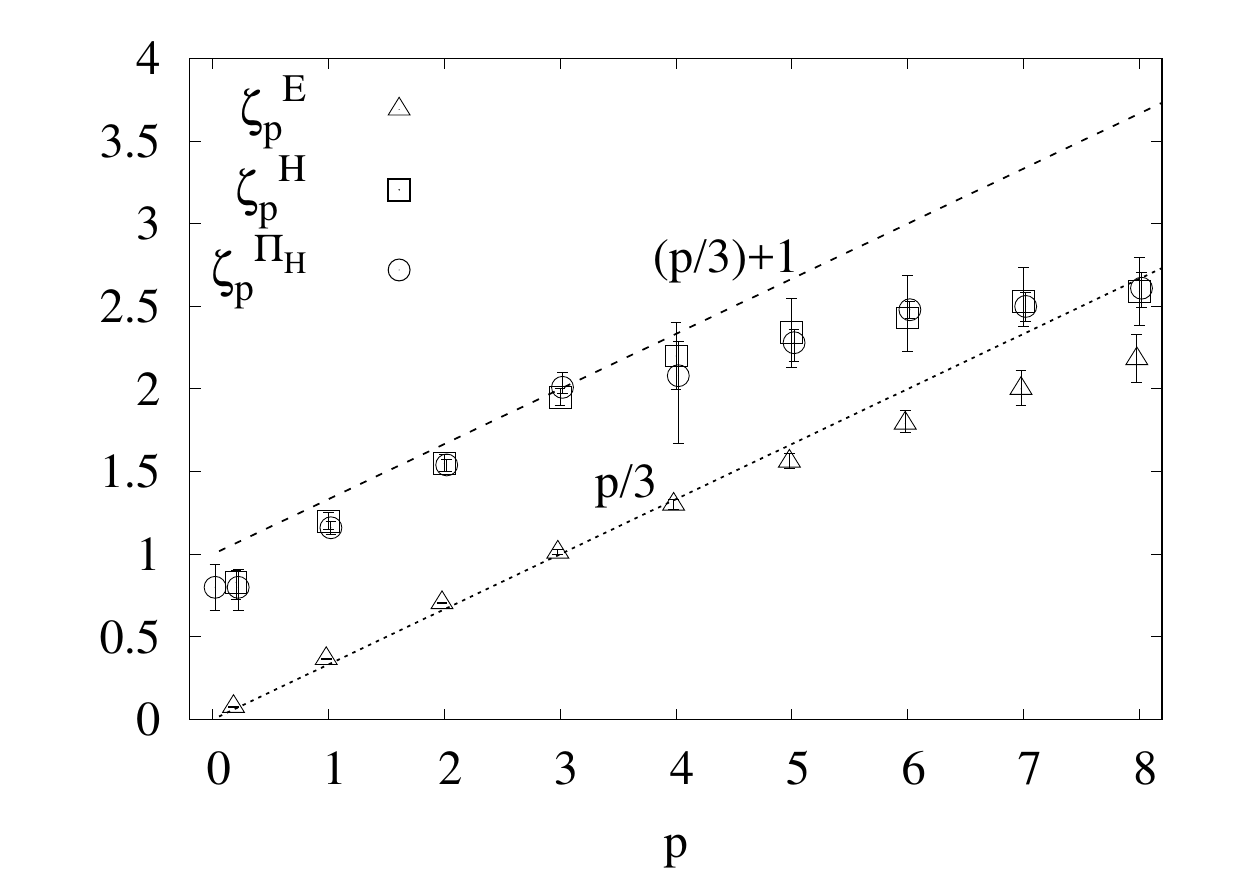}
\caption{Exponents of the symmetric and antisymmetric structure 
functions (\ref{eq:S_E})-(\ref{eq:S_PI_H}) as functions of order $p$. 
The values are obtained as the mean of linear fits over different ranges of 
shells.
The error bars show the dispersion as a function of the fitting ranges.}
\label{fig:9}
\end{figure} 
%%%%%%%%%%%%%%%%%%%%%%%%%%%%%%%%%%%%%%%%%%%%%%%%%%%%%%%%%%%%%%

%%%%%%%%%%%%%%%%%%%%%%%%%%%%%%%%%%%%%%%%%%%%%%%%%%%%%%%%%%%%%%
\begin {table}[htb!]
\caption{Helicity indices $s_i$ in (\ref{eq:sabra_helical_standard}) 
for the four helical shell models.}
\label{tab:helicity_indices} 
\begin{tabular*}{\linewidth}{@{\extracolsep{\fill} } c  c  c  c  c  c  c }
    \toprule
    Model & $s_1$ & $s_2$ & $s_3$ & $s_4$ & $s_5$ & $s_6$ \\ \colrule
    No. $1$   &  $+$ &  $-$ &  $-$ &  $-$ &  $-$ &  $+$ \\
    No. $2$   &  $-$ &  $-$ &  $+$ &  $-$ &  $+$ &  $-$ \\
    No. $3$   &  $-$ &  $+$ &  $-$ &  $+$ &  $-$ &  $-$ \\
    No. $4$   &  $+$ &  $+$ &  $+$ &  $+$ &  $+$ &  $+$ \\
    \botrule
\end{tabular*}
\end{table}
%%%%%%%%%%%%%%%%%%%%%%%%%%%%%%%%%%%%%%
%%%%%%%%%%%%%%%%%%%%%%%%%%%%%%%%%%%%%%
\begin {table}[htb!]
\caption{Coefficients of Eqs. (\ref{eq:sabra_helical_standard}) 
for the four helical shell models. These values depend on the 
shell-to-shell ratio $\lambda = k_n/k_{n-1}$; here $\lambda=2$. These 
coefficients guarantee energy and helicity conservation. 
Conventionally, and without loss of generality, we always choose 
$a=1$.}
\label{tab:model_coefficients} 
\begin{tabular*}{\linewidth}{@{\extracolsep{\fill} } c  c  c }
    \toprule
    Model & $b$ & $c$ \\ \colrule
    No. $1$&$-1/2$&$1/2$\\
    No. $2$&$-5/2$&$-3/2$\\
    No. $3$&$-5/6$&$1/6$\\
    No. $4$&$-3/2$&$-1/2$\\
    \botrule
\end{tabular*}
\end{table}
%%%%%%%%%%%%%%%%%%%%%%%%%%%%%%%%%%%%%%

The four classes of interactions conserve
energy and helicity separately, as in the original NSE, provided that the 
coefficients $a$ , $b$, and $c$ are
chosen appropriately. The added value with respect to simpler shell models is
that the energy and helicity now have structures similar to
(\ref{eq:Etot})--(\ref{eq:Htot}) for the NSE \cite{waleffe}: 
\begin{align}
\label{eq:energy_def}
E = E^+ + E^- & = \sum_{n=0}^N ( |u_n^+|^2 + |u_n^-|^2) \, ,\\
\label{eq:helicity_def}
H = H^+ + H^- & = \sum_{n=0}^N k_n( |u_n^+|^2 - |u_n^-|^2).
\end{align}
Here we consider only model no. $3$ because its dynamics is 
known to be dominated by a forward  energy cascade, with scaling 
properties very similar to those of the original NSE
\cite{biferale_helical_1996, chen2003joint}. The equations that we 
integrate are the following: 
\begin{equation}
\begin{aligned}
\label{eq:sabra_model3}
\dot{u}_n^+ =  i& (a k_{n+1} u_{n+2}^{-} u_{n+1}^{+*} + b k_{n} u_{n+1}^{-} 
u_{n-1}^{+*} \\
& + c k_{n-1} u_{n-1}^{-} u_{n-2}^{-}) - \nu k_n^2  u_n^+ + f_n^+  \, ,\\
\dot{u}_n^- = i& (a k_{n+1} u_{n+2}^{+} u_{n+1}^{-*} + b k_{n} u_{n+1}^{+} 
u_{n-1}^{-*} \\
& + c k_{n-1} u_{n-1}^{+} u_{n-2}^{+}) - \nu k_n^2   u_n^- + f_n^-  \, .
\end{aligned}
\end{equation}
We used a fully helical forcing, injecting energy only on the positive modes of
the first two shells $f^+_0=\xi_{r,0} + i \xi_{i,0}$, $f^+_1=0.5 (\xi_{r,1} + i
\xi_{i,1})$ (where all $\xi$ are Gaussian random variables with $\langle
\xi \rangle=0$ and $\langle \xi^2 \rangle=1$), in order to mimic the set-up of
the previous section. The number of shells is $N=25$, $k_0=1$, the
shell-to-shell ratio is $\lambda = k_n/k_{n-1}=2$, and the viscosity is $\nu=1.5
\cdot 10^{-7}$. The time integration is given by a second order Adams-Bashforth
scheme, with explicit integration of the viscous term
\cite{ultimo_articolo_shellmodels}.  With this setup, the Reynolds number is
$Re \sim 10^7$, and the large scale eddy turnover time is $\tau_0 \sim 1$.  We
let the system evolve for a total time $\sim 10^5 \tau_0$. In Fig.~\ref{fig:6}
we show the typical evolution of the total energy and of the total helicity in
one simulation.

In full analogy with the definitions used in Sec.  
\ref{sec:section2} for NSE, we can write mirror-symmetric and 
mirror-antisymmetric structure functions as:
\begin{align}
&{\cal S}^E_p(k_n) = \langle |u_n^+|^p \rangle + \langle |u_n^-|^p \rangle \sim 
k_n^{-\zeta_p^E} \, , \label{eq:S_E}\\
&{\cal S}^H_p(k_n) = \langle |u_n^+|^p \rangle - \langle |u_n^-|^p \rangle \sim 
k_n^{-\zeta_p^H} \, . \label{eq:S_H}
\end{align}
Additionally, we can define mirror-antisymmetric structure functions based on 
the third order correlation function responsible for the helicity flux 
\cite{ditlevsen2000, biferale_pierotti_helicity}:
\begin{equation}
\label{eq:S_PI_H}
{\cal S}^{\Pi_H}_p(k_n) =  \left\langle \sign( \Pi^H_n ) \left| k_n^{-2} \, 
\Pi^H_n  \right|^{p/3}  \right\rangle \sim k_n^{-\zeta_p^{\Pi_H}}  \, ,
\end{equation}
where for shell model (\ref{eq:sabra_model3}) the instantaneous 
helicity flux at shell $n$ is:
\begin{equation}
\label{eq:flux_hel_model3}
\Pi^H_n  = \sum_{i=0}^n \dot{H}_i =\left( \frac{a}{\lambda} + b \right) 
\delta^H_n +  \frac{a}{\lambda} \delta^H_{n+1} \, ,
\end{equation}
where $\delta^H_n = -2 k_{n}^{2} (C_{3,n}^+ - C_{3,n}^-)$ 
with $C_{3,n}^\pm = Im ( u_{n+1}^{\mp} u_{n}^{\pm*} u_{n-1}^{\pm*})$. 
Also for shell models it is possible to use a dimensional argument to predict 
the
scaling of mirror-symmetric and mirror-antisymmetric quantities. Let us 
consider the energy and helicity 
balance equations in the inertial range, where dissipative effects are 
negligible:
\begin{equation}
\varepsilon = \langle \Pi^E_n \rangle \, , \qquad h =  \langle \Pi^H_n \rangle 
\, ,
\end{equation}
where $\varepsilon$ and $h$ are the energy and helicity input at large scales, 
respectively. 
The instantaneous energy flux at shell $n$ is:
\begin{equation}
\label{eq:flux_en_model3}
\Pi^E_n  = \sum_{i=0}^n \dot{E}_i =\left( a + b \right) \delta^E_n +  a 
\delta^E_{n+1} \, ,
\end{equation}
where $\delta^E_n = -2 k_{n} (C_{3,n}^+ + C_{3,n}^-)$, 
while the helicity flux is defined in (\ref{eq:flux_hel_model3}). As a 
consequence, 
\begin{align}
\langle C_{3,n}^\pm \rangle & \sim \varepsilon k_n^{-1} \pm h k_n^{-2} \, . 
\end{align}
We can then identify
\begin{align}
\langle |u_n^\pm| \rangle \sim \langle |C_{3,n}^\pm|^{1/3} \rangle \, . 
\label{eq:sm_u_prediction}
\end{align}
Substituting (\ref{eq:sm_u_prediction}) in (\ref{eq:S_E}) and (\ref{eq:S_H}) 
and considering that 
(\ref{eq:S_PI_H}) should have the same chirality and dimensions of 
(\ref{eq:S_H}), we get the predictions:
\begin{align}
&{\cal S}^E_p(k_n) \sim k_n^{-p/3}\, , \label{eq:shell_structure_prediction_E}\\
&{\cal S}^H_p(k_n) \sim {\cal S}^{\Pi_H}_p(k_n) \sim k_n^{-(p/3+1)}\, . 
\label{eq:shell_structure_prediction_H}
\end{align}

In Fig.~\ref{fig:7} we show the scaling observed for ${\cal
S}^E_p(k_n)$, ${\cal S}^H_p(k_n)$ and ${\cal S}^{\Pi_H}_p(k_n)$ at changing
$p$. The scaling regime for the helical components is
deteriorating for higher moments.  In particular we  observe a change of sign 
for
the antisymmetric structure functions in the middle of the inertial range, 
hence 
we plot the absolute values. Spurious contributions to the
powerlaw scaling can be a consequence of contaminations coming from the viscous
range or from inertial subleading terms.  In order to clarify this
point, we performed another set of simulations with
$N=31$ shells and $Re \sim 3 \cdot 10^9$. As can be seen from Fig.~\ref{fig:8},
even with a longer inertial range, the change of sign and
the deterioration in the scaling for high-order helical structure functions are
still present, indicating that viscosity might not be the
primary cause.  In Fig.~\ref{fig:9} we summarize the behavior of all scaling
exponents, compared with the dimensional predictions
(\ref{eq:shell_structure_prediction_E})--(\ref{eq:shell_structure_prediction_H}).
For higher orders, a deviation from the dimensional prediction is
observed. As seen in Sec. \ref{section:NS} for the original NSE, this is 
possibly due to  
intermittent corrections or subleading contributions coming
from sub-leading corrections in the helicity flux.  
Even though the presence of a change of sign in higher-order moments results in
large errorbars in the estimate of the scaling exponents, our 
measurements are in good agreement with those reported in
\cite{ditlevsen2000}. Our observations disagree with the
scaling of  subgrid helicity flux measured in DNS of the NSE, reported in Ref. 
\cite{chen2003prl}. However, in the latter case, the  helicity flux is taken 
with absolute values, leading to 
a possible mixing among chiral and mirror-symmetric contributions.  

\section{Conclusions}

We have studied the statistical properties of helicity in DNS of high
Reynolds number flows. We focused on a set of observables sensitive to
mirror symmetry, studying the scaling properties of structure functions based
on either helical-projected velocity fields or on velocity-vorticity
correlations.  In both cases we found that chiral contributions are subleading
with respect to their counterparts involving only mirror-symmetric components.
We investigated the scaling behavior of these subleading corrections in
high-order structure functions. A dimensional argument assuming that the main
chiral contributions are analytical in the helicity flux captures the power law
scaling quite well, except for some anomalous correction. Controlling the
multiscale amplitudes of chiral fluctuations is key to develop also subgrid
turbulent models for flows that break mirror symmetry either globally or
locally \cite{yu2014}.  Furthermore, we extended our analysis to higher
Reynolds numbers by measuring the statistics of helicity in shell models. Also
in shell models we found a scaling behavior quantitatively similar to what
reported for the Navier-Stokes equations, including the presence of correction
to scaling even at extremely high Reynolds numbers.

\section*{Acknowledgments}
We thank M. Sbragaglia useful discussions. The research leading to these results has received funding from  the European
Union's Seventh Framework Programme (FP7/2007-2013) under ERC grant agreement 
No.
339032 and from COST Action MP1305.
\bibliography{mybib}

\end{document}